\documentclass[%
 reprint,
 superscriptaddress,
 amsmath,amssymb,
 aps, prb
]{revtex4-2}

\usepackage{graphicx}
\usepackage{float}
\usepackage{dcolumn}
\usepackage{bm}
\usepackage{siunitx}
\usepackage{xcolor}
\usepackage{booktabs}
\usepackage{natbib, hyperref}
\hypersetup{
	colorlinks=true,
	linkcolor=blue,      
	urlcolor=blue,
	citecolor=blue
}

\begin{document}

\preprint{APS/123-QED}

\title{The interplay between high-harmonic generation and photoluminescence in ZnO: Anisotropic spectral properties of harmonic emission and the role of excitons}

\author{Simon P. S. Jessen}
\affiliation{Aarhus University, Department of Physics and Astronomy, Ny Munkegade 120, 8000 Aarhus C, Denmark}

\author{Mark Mero}
\affiliation{
 Max-Born-Institute for Nonlinear Optics and Short Pulse Spectroscopy, Max-Born-Strasse 2A, D-12489 Berlin, Germany}

\author{Marc J. J. Vrakking}
\affiliation{
 Max-Born-Institute for Nonlinear Optics and Short Pulse Spectroscopy, Max-Born-Strasse 2A, D-12489 Berlin, Germany}

\author{Rosana M. Turtos}
\affiliation{Aarhus University, Department of Physics and Astronomy, Ny Munkegade 120, 8000 Aarhus C, Denmark}

\author{Peter Balling}
\affiliation{Aarhus University, Department of Physics and Astronomy, Ny Munkegade 120, 8000 Aarhus C, Denmark}

\author{Peter Jürgens}
\email{juergens@mbi-berlin.de}
\affiliation{
 Max-Born-Institute for Nonlinear Optics and Short Pulse Spectroscopy, Max-Born-Strasse 2A, D-12489 Berlin, Germany}

\date{\today}


\begin{abstract}
We investigate the nonlinear optical response of bulk ZnO under intense short-wave infrared excitation, focusing on the interplay between high-harmonic generation (HHG) and photoluminescence (PL). While HHG exhibits non-perturbative intensity scaling and a spectral blueshift consistent with plasma-induced refractive index changes, the PL signal shows a pronounced superlinear increase and a redshift, attributed to a combination of exciton–exciton scattering and phonon-assisted exciton recombination emission. A similar PL response under above-bandgap excitation supports its intrinsic origin. Spectral analysis of the HHG emission reveals an intensity-driven transition in the characteristics of the fifth harmonic, indicating a change in the underlying generation mechanism. These findings establish PL and spectral HHG analysis as complementary probes of strong-field and many-body effects in wide-bandgap semiconductors.

\end{abstract}

\maketitle
\section{Introduction}
Since its first experimental demonstration~\cite{Chin_2001,Ghimire_2011}, high-harmonic generation (HHG) in solids has rapidly evolved into a distinct and expanding research field. It has driven major advances in attosecond physics in condensed matter systems~\cite{Li_2020}, petahertz optoelectronics~\cite{Ossiander_2022, Heide_2024}, and ultrafast photoconductive sampling~\cite{Zimin_2021, Sederberg_2020}. Over the past decade, numerous experimental and theoretical studies of HHG in bulk crystals and quantum materials have opened up new directions, including band-structure reconstruction~\cite{Luu_2015, Lanin_2017}, analysis of sub-cycle carrier dynamics~\cite{Zaks_2012}, Floquet engineering~\cite{Uchida_2024}, and probing of electron-hole coherence times~\cite{Heide_2022}. 

Among the materials extensively investigated for solid-state HHG, ZnO has emerged as a prototypical system due to its wide bandgap and strong nonlinear optical response. Several key studies have explored HHG in ZnO~\cite{Gholam_2018}, employing various experimental configurations such as polarization-resolved~\cite{Hollinger_2020}, crystal-orientation-resolved~\cite{Ghimire_2011, Jiang_2019, Gholam_2017}, wavelength-dependent~\cite{Gholam_2017}, and spatially-resolved measurements~\cite{Li_2022}. These static HHG experiments were later complemented by time-resolved studies that examined the influence of photocarrier doping on the HHG yield and the spectral characteristics~\cite{Wang_2017, Xu_2022, Nie_2024}. 

Interestingly, many of these experiments have also reported a strong emission near \SI{390}{\nano\meter}, commonly attributed to photoluminescence (PL) from radiative electron–hole recombination or defect-assisted excitonic decay~\cite{Hollinger_2020}. While often treated as a secondary effect, this PL emission may carry complementary information about strong-field carrier excitation and relaxation dynamics. A few studies have started to explore this potential, using the PL signal as an indicator for excitation strength or ionization probability~\cite{Hollinger_2020, Liu_2021, Truong_2025}. In particular, the PL yield is increasingly employed in field-resolved metrology as a nonlinear signal for the sampling of ultrashort mid-infrared and near-infrared laser pulses. However, a reliable and reproducible characterization requires a deeper understanding of the physical origin of this signal and its dependence on both laser and material parameters.

In this work, we investigate the correlation between high-harmonic emission and the accompanying PL signal from bulk wurtzite-type ZnO crystals excited by intense, ultrashort short-wave infrared (SWIR) laser pulses. In addition to the strong-field regime, we also explore the PL emission under above-bandgap excitation. We attribute the PL to radiative recombination of excitonic states, mediated by either phonon-assisted recombination or exciton–exciton scattering, resulting in a luminescence band that is considerably (i.e.~\SI{200}{\milli\electronvolt}) redshifted from the band edge of the material. Our results reveal clear differences in the intensity scaling and spectral characteristics of the coherent HHG emission and the incoherent PL signal. Together, these findings offer new insight into strong-field light–matter interaction in wide-bandgap semiconductors and the nonlinear optical response of ZnO under extreme excitation conditions.


\section{Experimental Setup}

\begin{figure}[ht]
    \centering\includegraphics[width=\linewidth]{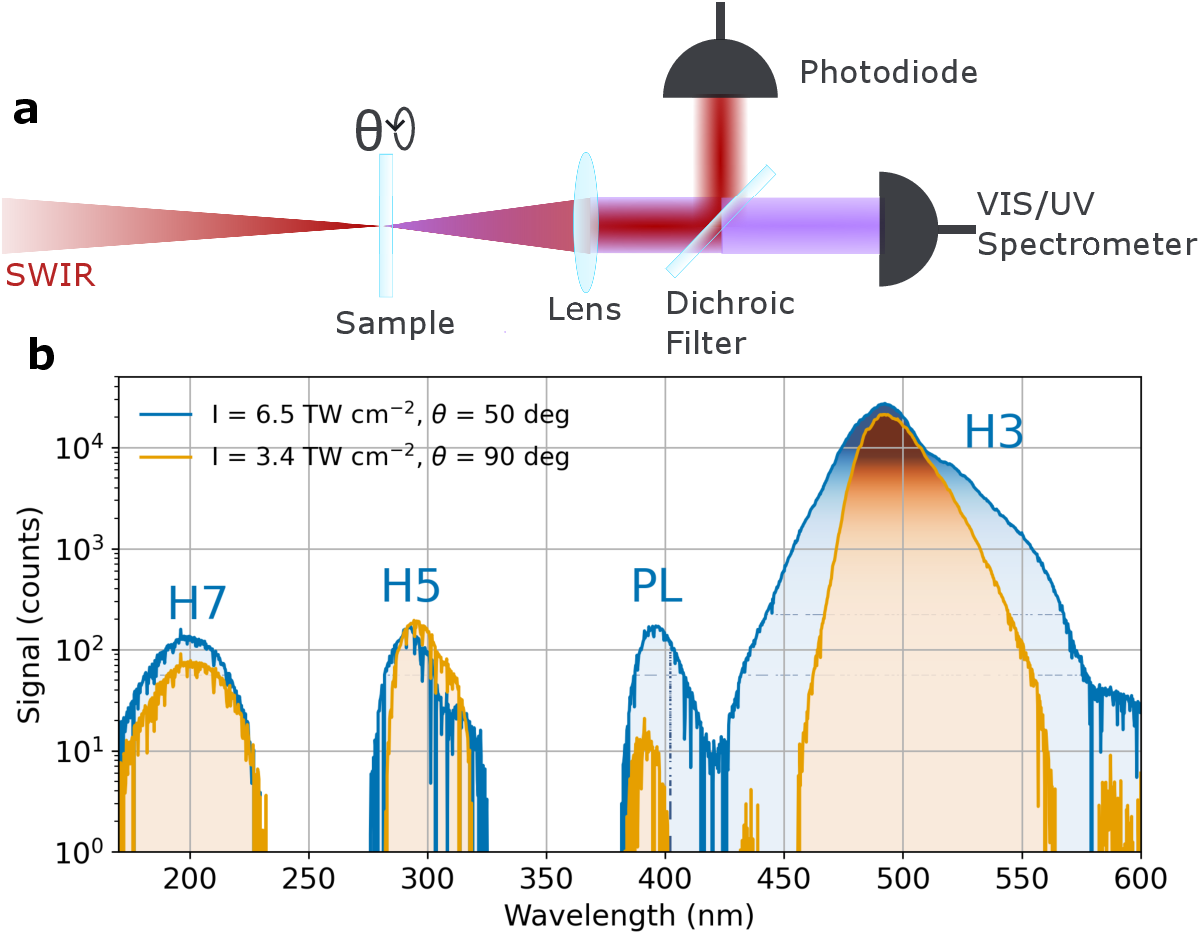}
    \caption{Experimental setup and exemplary result. (a) Setup for the detection of high-harmonics and photoluminescence from ZnO.  (b) Spectrum containing odd harmonics of the fundamental SWIR radiation up to the seventh order and the PL signal, at SWIR peak intensities of \SI{6.5}{\tera\watt\per\centi\meter\squared} (blue) and \SI{3.4}{\tera\watt\per\centi\meter\squared}  (orange).}
\label{fig:fig_1}
\end{figure}

The experimental setup is illustrated in Fig.~\ref{fig:fig_1}(a). A home-built, dual-beam optical-parametric chirped-pulse amplification system, similar to that described in Ref.~\cite{Juergens_2024}, was used to generate driving laser pulses at a wavelength of \SI{1500}{\nano\meter} with a full-width half-maximum pulse duration of \SI{50}{\femto\second}. The beam was focused into the bulk of a \SI{200}{\micro\meter}-thick single-crystalline wurtzite-type ZnO crystal with a (0001)-oriented surface using an off-axis parabolic mirror with a focal length of \SI{25}{\milli\meter}. The generated high-harmonic radiation and PL emission were spectrally analyzed using a commercial fiber-based spectrometer (Avantes AvaSpec-HS1024x58/122TEC). 

The ZnO sample was mounted on a computer-controlled rotation stage, allowing precise adjustment of the relative angle $\theta$ between the incident linear laser polarization and the crystallographic c-axis of the sample. This enabled systematic studies of the orientation-dependent emission characteristics. 

High-harmonic emission up to the seventh order of the excitation frequency was detected as shown in Fig.~\ref{fig:fig_1}(b). In addition to the harmonic signals, a strong emission peak around \SI{390}{\nano\meter} was observed, consistent with previous reports \cite{Ghimire_2011, Hollinger_2020, Wang_2017}. Figure \ref{fig:fig_1}(b) presents representative spectra recorded at SWIR peak intensities of \SI{6.5}{\tera\watt\per\centi\meter\squared} [blue line in Fig.~\ref{fig:fig_1}(b)] and \SI{3.4}{\tera\watt\per\centi\meter\squared} [orange line in Fig.~\ref{fig:fig_1}(b)], clearly displaying the third, fifth and seventh harmonic orders along with the PL emission. As discussed in detail in Ref.~\cite{Juergens_2024}, we simultaneously measured the transmission of the SWIR pump beam through the strongly excited ZnO sample. 
%
%

To complement the SWIR excitation measurements, we also investigated the PL response of ZnO under above-bandgap (\SI{6.05}{\electronvolt}) ultrafast (\SI{\sim80}{\femto\second}) optical excitation. In order to reveal fluence-dependent variations in the PL efficiency arising from radiative and non-radiative recombination processes involving multiple emissive species, these measurements were performed using the z-scan luminescence method \cite{Grim_2013}, in which the PL is recorded under constant excitation energy but varying excitation density, achieved by translating a lens that focuses the fourth harmonic of a Ti:sapphire laser onto the crystal. With a fixed pulse energy of \SI{42}{\nano\joule} and variable beam width, peak fluence levels between \SI{0.01}{\milli\joule\per\square\centi\meter} and \SI{2}{\milli\joule\per\square\centi\meter} were accessed. This technique is designed to  A detailed description of the specific z-scan setup used in this work can be found in Ref.~\cite{Jessen_2025}.

\section{Experimental Results}
\begin{figure}[ht]
    \centering\includegraphics[width=0.9\linewidth]{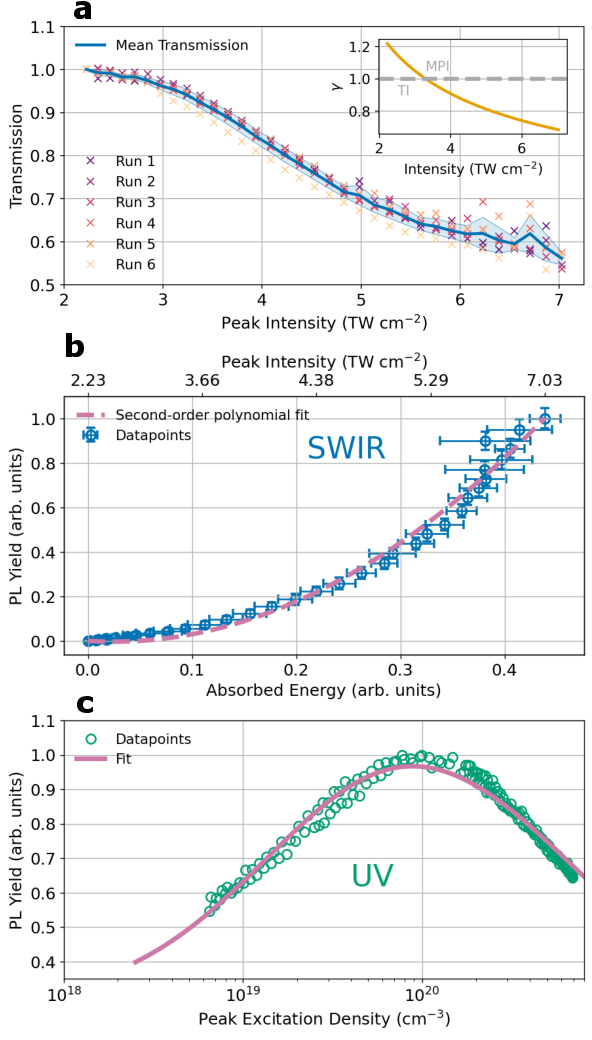}
    \caption{Experimentally determined intensity dependence of the SWIR-transmission and PL yield. (a) Transmission of the SWIR laser pulse through the ZnO crystal (averaged over all crystal orientations and six consecutive measurements) as a function of the SWIR peak intensity. The inset shows the corresponding value of the Keldysh parameter $\gamma$. (b) Experimentally determined yield of the PL signal induced by SWIR laser excitation as a function of the absorbed energy extracted from (a) (bottom axis) and the SWIR peak intensity (top axis). (c) Experimentally measured PL yield as a function of excitation density under above-bandgap excitation. The excitation density is estimated using an absorption coefficient of ZnO of \SI{4.2e5}{\per\centi\meter}~\cite{Seitz_1959} and a reflectance of 0.1~\cite{Benkrima_2023}.}
    \label{fig:fig_2}
\end{figure}

%
Figure~\ref{fig:fig_2}(a) shows the measured transmission of the SWIR pump laser beam as a function of excitation intensity, averaged over all crystal orientations and six consecutive measurements. As in previous nonlinear transmission studies \cite{Grojo_2013, Sneftrup_2024}, the transmission remains constant at low intensities, then decreases nonlinearly and saturates near the threshold for irreversible material modification. This reduction is attributed to absorption by the dense electron-hole plasma, with the onset of transmission loss marking the onset of strong-field excitation. 

The inset in Fig.~\ref{fig:fig_2}(a) shows the corresponding Keldysh-parameter $\gamma$, defined as 
\begin{equation}
    \label{eq:keldysh}
    \gamma = \omega_0 \left( \frac{m^{\ast} c \epsilon_0 n_0 E_g}{e^2 I} \right)^{1/2},
\end{equation}
where $\omega_0$ is the central laser frequency, $e$ the elementary charge, $m^{\ast} = 0.24 m_{e}$ the effective electron mass, $c$ the speed of light in vacuum, $n_0$ the linear refractive index, $\epsilon_0$ the vacuum permittivity, $E_g$ the bandgap energy, and $I$ the peak laser intensity. A value of $\gamma \geq 1$ indicates multiphoton-dominated excitation, while $\gamma < 1$ corresponds to a tunneling regime. 

Assuming that the electron-hole densities generated below the damage threshold are too low to induce significant reflectivity changes, and that reflection remains negligible, the transmission measurements in Fig.~\ref{fig:fig_2}(a) can be used to estimate the fraction of energy absorbed in the ZnO crystal. This calibration allows us to plot the PL yield near \SI{390}{\nano\meter} as a function of absorbed SWIR energy [Fig.~\ref{fig:fig_2}(b)], revealing a monotonic increase that is well-described by a second-order polynomial fit (see purple line). As an additional estimate, we compute the peak carrier density by dividing the absorbed energy by the bandgap energy and assuming a cylindrical excitation volume, yielding densities on the order of \SI{1e20}{\per\cubic\centi\meter}. This value is used to compare the excitation conditions to those used in above-bandgap excitation experiments dscussed below.
%

A comparable superlinear increase in PL yields is also observed in the interband z-scan luminescence measurements shown in Fig.~\ref{fig:fig_2}(c). The peak fluence is converted to an estimated excitation density (in units of electron-hole pairs per volume) using the absorption coefficient of ZnO at \SI{6}{\electronvolt}, taken as \SI{4.2e5}{\per\centi\meter} \cite{Seitz_1959}, and assuming a reflectance of 0.1~\cite{Benkrima_2023}. Since the total optical excitation energy remains constant during the z-scan, an increasing or decreasing trend in the detected signal predicts regimes of excitation density with a superlinear or sublinear intensity response, respectively. 
%

The resulting non-monotonic PL response is well captured by a model that adds a second-order radiative term to the standard rate-equation formalism commonly used for self-activated scintillators and wide-bandgap semiconductors \cite{Kirm_2009, Grim_2013, Spassky_2019} [see purple line in Fig.~\ref{fig:fig_2}(c)]. A detailed description of the numerical model is provided in Sec.~IV.

\begin{figure}[ht]
    \centering\includegraphics[width=\linewidth]{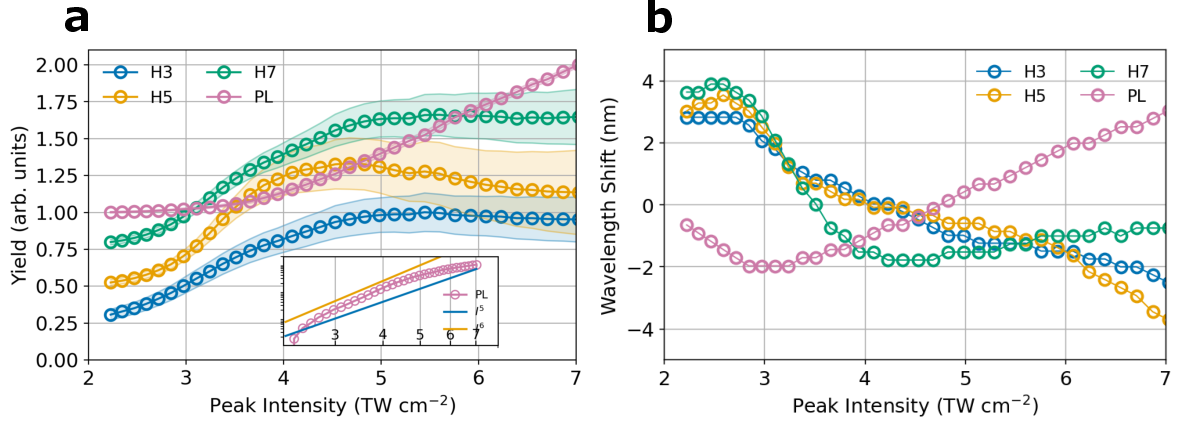}
    \caption{Experimental investigation of the observed high-harmonic and PL emission from a bulk ZnO crystal. (a) Dependence of the harmonic signal yields and the PL signal on the excitation intensity of the SWIR pump laser pulse. The data have been vertically offset for clarity by 0.33 for H5, 0.66 for H7, and 1.0 for the PL signal. (b) Variation of the central wavelength of the observed signals as a function of the SWIR pump laser intensity.}
    \label{fig:fig_3}
\end{figure}

The measured signal strengths of the odd harmonics and the PL emission as a function of the SWIR peak intensity are presented in Fig.~\ref{fig:fig_3}(a). At moderate peak intensities ($\leq\,$\SI{4}{\tera\watt\per\centi\meter\squared}), all observed harmonic orders exhibit a strong nonlinear increase with increasing laser intensity. As the intensity approaches the damage threshold of the ZnO crystal (at $\sim\,$\SI{8}{\tera\watt\per\centi\meter\squared}), the harmonic signal begins to saturate - a characteristic feature of non-perturbative HHG processes - consistent with previous findings in ZnO \cite{Gholam_2018, Ghimire_2011} and other bulk crystals. 

In contrast, the intensity scaling of the PL emission follows a different trend, increasing nonlinearly across the entire intensity range, with no clear saturation observed. The inset of Fig.~\ref{fig:fig_3}(a) compares the PL yield to fifth- and sixth-order power law dependencies, showing that in the intermediate intensity range (\SIrange{4}{5}{\tera\watt\per\centi\meter\squared}), the signal follows an $I^6$ scaling before the nonlinearity decreases near the damage threshold. 
Note that, in contrast to Fig.~\ref{fig:fig_2}(b), where the PL signal is shown as a function of absorbed energy to enable comparison to carrier densities extracted under above-bandgap excitation, here we focus on the dependence of the PL yield on the incident SWIR intensity to directly compare its scaling behavior to that of the high-harmonic signals.

%
While the intensity scaling of signal yields provides insight into the underlying nonlinear processes, particularly the transition from perturbative to non-perturbative regimes, variations in the spectral properties are often overlooked in HHG studies, despite their potential to reveal important information about the frequency conversion mechanisms. Figure~\ref{fig:fig_3}(b) shows the shift in the central wavelengths of the observed odd harmonics and the PL emission as a function of the SWIR peak intensity. The central wavelength of each spectral feature is determined from the spectral center of mass, which offers greater robustness against asymmetric line shapes compared to fitting with symmetric functions such as Gaussians or Lorentzians. The plotted values represent the deviation of each central wavelength from its mean value across all excitation intensities. 

All harmonic orders exhibit a clear blueshift with increasing intensity, whereas the PL emission shows a redshift of up to \SI{4}{\nano\meter} at the highest intensities. Notably, the third and fifth harmonics display comparable scaling of the blueshift, while the seventh harmonic shows a steeper initial shift followed by saturation.

\begin{figure}[ht]
    \centering\includegraphics[width=\linewidth]{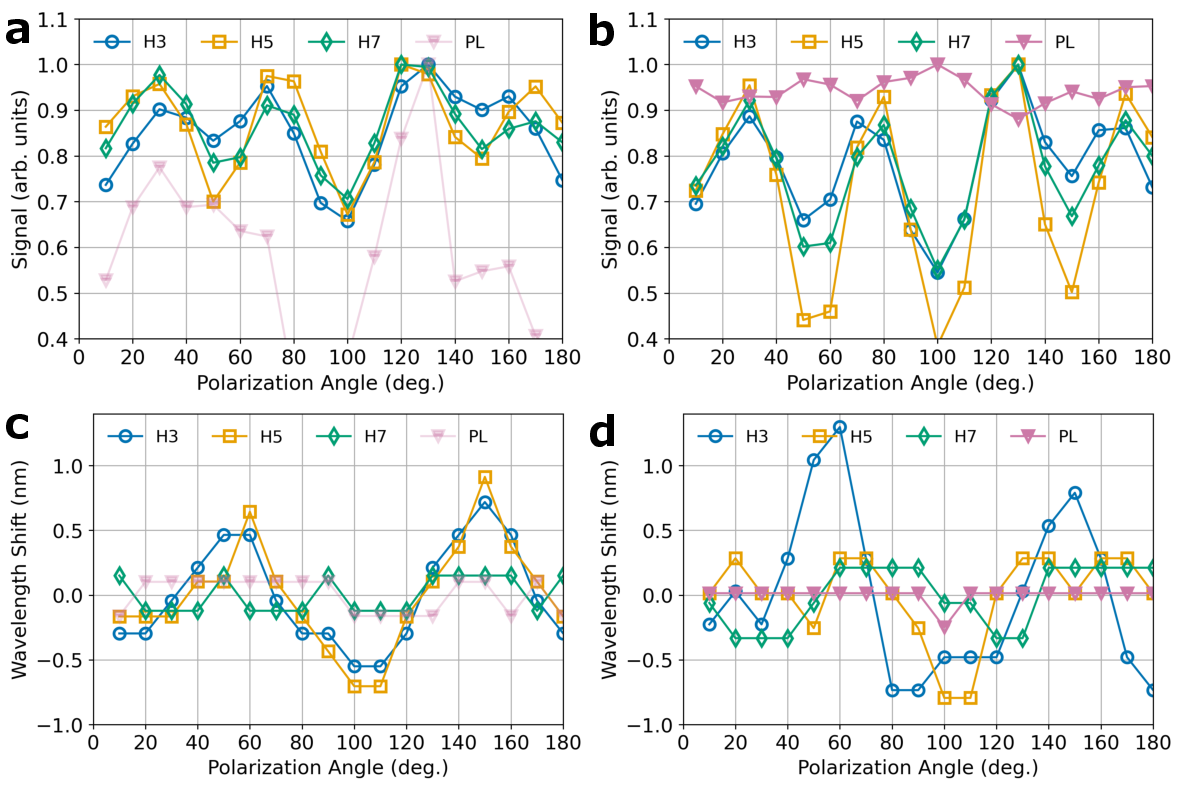}
        \caption{Orientation dependent measurements of HHG and PL in ZnO. (a). Angular dependence of the signal yields at a SWIR peak intensity of \SI{2.8}{\tera\watt\per\centi\meter\squared}. (b). Same as in (a) at a peak intensity of \SI{4.7}{\tera\watt\per\centi\meter\squared}. (c) Variation of the central wavelength as a function of the orientation angle $\theta$ at a SWIR peak intensity of \SI{2.8}{\tera\watt\per\centi\meter\squared}. (d) Same as in (c) at a peak intensity of \SI{4.7}{\tera\watt\per\centi\meter\squared}.}
    \label{fig:fig_4}
\end{figure}

%
To further investigate the influence of band structure and crystal orientation, we measured the dependence of both HHG and PL signals on the angle $\theta$ between the linear laser polarization and the principal crystallographic axis of the ZnO crystal. Figures~\ref{fig:fig_4}(a) and (b) show the intensity dependence of the odd harmonics and PL emission as a function of $\theta$ at SWIR peak intensities of \SI{2.8}{\tera\watt\per\centi\meter\squared} and \SI{4.7}{\tera\watt\per\centi\meter\squared}, respectively. The harmonic signals exhibit a pronounced fourfold symmetry, while the PL response remains isotropic, consistent with previous observations~\cite{Ghimire_2011}. At low intensity in Fig.~\ref{fig:fig_4}(a) the PL signal is in the range of the detection limit, hence no pronounced, reproducible angular dependence can be extracted. The observed $\pi$-periodicity in the harmonic yields agrees with earlier experimental and theoretical studies~\cite{Ghimire_2011, Jiang_2019}.  

Figures~\ref{fig:fig_4}(c) and (d) present the angular dependence of the central wavelength of the harmonics and PL emission. At a moderate SWIR intensity of \SI{2.8}{\tera\watt\per\centi\meter\squared}, the third and fifth harmonics exhibit two distinct maxima in their wavelength shifts, while the seventh harmonic and PL emission show no significant angular dependence. At higher intensity (\SI{4.7}{\tera\watt\per\centi\meter\squared}), only the third harmonic retains an anisotropic spectral response with increased amplitude, whereas the fifth harmonic becomes nearly isotropic, similar to the seventh harmonic and the PL signal, whose apparent invariance is limited by the spectrometer resolution of \SI{0.26}{\nano\meter}.

%


\section{Discussion}
To elucidate the microscopic origin of the PL signal observed near \SI{390}{\nano\meter}, we revisit its assignment in the context of strong-field excitation in ZnO. While this emission has often been labeled as "bandgap fluorescence" in the solid-state HHG literature, its spectral position and nonlinear excitation behavior suggest a more specific mechanism. At room temperature, direct free exciton recombination in ZnO is expected to occur near \SI{375}{\nano\meter}, based on a bandgap of \SI{3.37}{\electronvolt} and an exciton binding energy of $\sim\,$\SI{60}{\milli\electronvolt} \cite{Ozgur_2005, Alivov_2003}. The observed emission, however, is consistently redshifted by $\geq\,$\SI{15}{\nano\meter}, which excludes direct band-to-band or free exciton recombination as the dominant origin. However, as previously reported \cite{Klingshirn_2012}, the phonon replicas of free exciton emission can still contribute to the observed spectra.  

Another exciton-based mechanism that would fit the observed PL spectra is that of collisional exciton recombination, also known as P-band luminescence \cite{Grim_2013, Klingshirn_2012}. In this process, two excitons undergo inelastic scattering, with one recombining radiatively while the other is promoted to a higher energy state. The emitted photon is redshifted relative to the free exciton, consistent with the observed spectral position. This two-exciton process naturally gives rise to a quadratic dependence of the PL yield on the carrier density. While originally identified at cryogenic temperatures, collisional exciton emission has also been observed in high-quality ZnO films at room temperature in the absence of quantum confinement \cite{Chen_2001}. These findings establish the P-band as an intrinsic feature of ZnO's excitonic landscape, particularly at intermediate excitation densities where many-body interactions are prominent but exciton ionization is not yet dominant. 

Our experimental data provide strong support for this assignment. Under both SWIR and above-bandgap excitation, the PL yield increases superlinearly with absorbed energy / carrier density [see Figs.~\ref{fig:fig_2}(b,c)]. As such behavior deviates significantly from the linear or sublinear scaling typical of defect-related luminescence or electron-hole radiative recombination. Importantly, time-resolved studies have shown that exciton formation and carrier thermalization in ZnO occur on sub-picosecond timescales \cite{Baxter_2006, Deinert_2014, Milne_2023}, allowing sufficient time for exciton-exciton scattering even under ultrafast excitation. Taken together, the observed spectral redshift, the superlinear (quadratic) scaling, and the consistency across excitation regimes strongly support the assignment of the PL to phonon-mediated excitonic recombination and room-temperature collisional exciton emission. 

This interpretation is reinforced by further numerical analysis of the interband z-scan luminescence measurements, which show a non-monotonic PL response as a function of excitation density [Fig.~\ref{fig:fig_2}(c)]. Such a behavior is unusual among scintillators, which exhibit quenching at high fluences due to exciton-exciton annihilation or Meitner-Auger recombination \cite{Kirm_2009, Grim_2013, Spassky_2019, Wei_2021}. The non-monotonic response of the PL signal to increasing excitation densities is evidence of a radiative term with a higher order dependence on excitation density than the dominant quenching term.
%

To describe this behavior quantitatively, we apply a phenomenological kinetic model that includes both radiative and non-radiative channels up to second order and that has been successfully applied previously in Refs.~\cite{Grim_2013, Kirm_2009, Spassky_2019, Wei_2021}:
\begin{equation}
    \label{eq:simons_eq}
    \frac{dn}{dt} = - (R_1 + Q_1)n - (R_2 + Q_2) n^2
\end{equation}
where $n$ is the exciton density and $R_1$, $R_2$ and $Q_1$, $Q_2$ are first- and second-order radiative and quenching coefficients, respectively. Due to the fast formation of excitons, we neglect an explicit generation term and assume an initial exciton population proportional to the absorbed energy. The fit shown in Fig.~\ref{fig:fig_2}(c) reproduces the data well. At low excitation densities, the PL yield rises due to the second-order radiative term, $R_2 n^2$, associated with exciton-exciton scattering. At higher densities, the second-order quenching term, $Q_2 n^2$, becomes significant, leading to saturation and eventual reduction in PL. Notably, this saturation is absent under SWIR excitation, likely because the corresponding excitation densities do not reach the quenching regime here seen above excitation densities of \SI{1e20}{\per\cubic\centi\meter}. 

The second-order radiative term in the model provides a natural link to the collisional exciton emission mechanism discussed above. Inelastic exciton-exciton scattering populates emissive states within the P-band, giving rise to the observed quadratic scaling \cite{Chen_2001, Klingshirn_2012} at high intensities. Thus, the same microscopic process explains both the fluence-dependent PL yield and its emission characteristics across different excitation regimes. 

It is important to note the limitations of the seond-order radiative model presented in Fig.~\ref{fig:fig_2}(c). This model, originally developed for stationary excitons in wide-bandgap scintillators, does not account for exciton mobility in ZnO or the complex dynamics of exciton-exciton interactions under our excitation conditions. As such, the extracted radiative and quenching rates should be interpreted with caution and not regarded as quantitatively accurate. Nonetheless, the model illustrates that the inclusion of a second-order radiative recombination channel is sufficient to capture the observed nonlinear behavior qualitatively. More detailed studies are required to develop a physically comprehensive model for the decay kinetics in ZnO following above-bandgap excitation.

%

The intensity scaling of the harmonics yields follows a strongly nonlinear, non-perturbative dependence on the peak electric field of the SWIR driving laser, consistent with the nature of HHG. In contrast, the PL signal exhibits a different kind of nonlinearity: it increases monotonically across the full range of accessible SWIR intensities, following a smooth supralinear trend. Recent studies have suggested that PL emission under strong-field excitation can follow a power-law dependence of the form $I^q$ where $q$ reflects the effective multiphoton order of the underlying excitation process \cite{Liu_2021, Truong_2025}. In our measurements, we find that a sixth-order power law provides the best fit to the experimental PL yield, as shown in Fig.~\ref{fig:fig_3}(a). Considering the fundamental photon energy of \SI{0.83}{\electronvolt} (corresponding to a central wavelength of \SI{1500}{\nano\meter}) and the bandgap of ZnO (\SI{3.37}{\electronvolt}), a five-photon absorption process ($q=5$) would be the minimal requirement for interband excitation. However, in the intensity regime explored here, the optical field is sufficiently strong to induce notable modifications of the band structure via the dynamic Stark effect. This effect leads to a field-induced increase of the effective bandgap, which can e.g. be expressed in terms of the Keldysh parameter $\gamma$ [see Eg.~\ref{eq:keldysh}] as \cite{sergaeva2018ultrafast}
\begin{equation}
    \label{eq:bandgap_shift}
    \widetilde{E}_g = E_g \left(1 + \frac{1}{2 \gamma^2} \right) ,    
\end{equation}
which shifts the excitation threshold to higher photon orders and thereby modifies the apparent scaling exponent. Including this Stark-induced bandgap modification yields a corrected multiphoton order of $q=6$ across the relevant intensity range. Thus, the observed PL scaling reflects not only the intrinsic multiphoton excitation pathway, but also the impact of transient band structure dressing by the strong driving field.  
%
%

Spectral shifts, as shown in Fig.~\ref{fig:fig_3}(b), can arise from several mechanisms, including dynamic changes in the linear and nonlinear refractive index due to the optical Kerr effect (e.g., self-phase and cross-phase modulation) or from carrier excitation. Additionally, strong-field-induced modifications of the electronic band structure can alter the refractive index and influence the accumulated phase during propagation, leading to frequency shifts. 

The observed blueshift of the odd harmonics is commonly attributed to excitation-induced refractive index changes. As an electron-hole plasma forms, the refractive index decreases, resulting in increased phase accumulation and a corresponding shift toward higher photon energies \cite{verhoef_2008, van_2023, Koll_2025}. This mechanism provides a consistent explanation for the general intensity dependence of the wavelength shifts observed in the HHG spectra. 

In contrast, the PL emission exhibits a fundamentally different behavior. Its nearly linear redshift with increasing SWIR laser intensity suggests that neither plasma-induced refractive index changes nor coherent, field-driven effects such as Kerr-type optical nonlinearities or AC stark shifts are the dominant contributors. Instead, the PL shift reflects an intrinsic material response, pointing to a distinct physical origin that persists until the time of the PL emission. 

Under the assignment of the PL signal to excitonic emission, the central wavelength near \SI{390}{\nano\meter} serves as a direct probe of the exciton binding energy. The observed redshift, which progresses with increasing excitation density, is indicative of bandgap renormalization driven by carrier-carrier interaction. Similar behavior has been reported in ZnO \cite{Alivov_2003, Reynolds_2000, Dai_2014}, as well as in a variety of other systems, including conventional semiconductors, transition-metal dichalcogenides, and perovskite quantum dots \cite{Ziaja_2015, Liu_2019, Ren_2022}. The measured redshift of $\sim$\SI{4}{\nano\meter}, corresponding to an energy shift of \SI{32}{\milli\electronvolt}, is relatively large compared to values reported in continuous-wave or weak excitation regimes but remains consistent with strong excitation by ultrashort laser pulses near the damage threshold. 

%
%

%
The excitonic interpretation for the PL emission also explains the isotropic nature of the PL emission, in contrast to the pronounced anisotropy observed in both the signal yield and central wavelength of the high-harmonic emission. \\

\begin{figure}[ht]
    \centering\includegraphics[width=\linewidth]{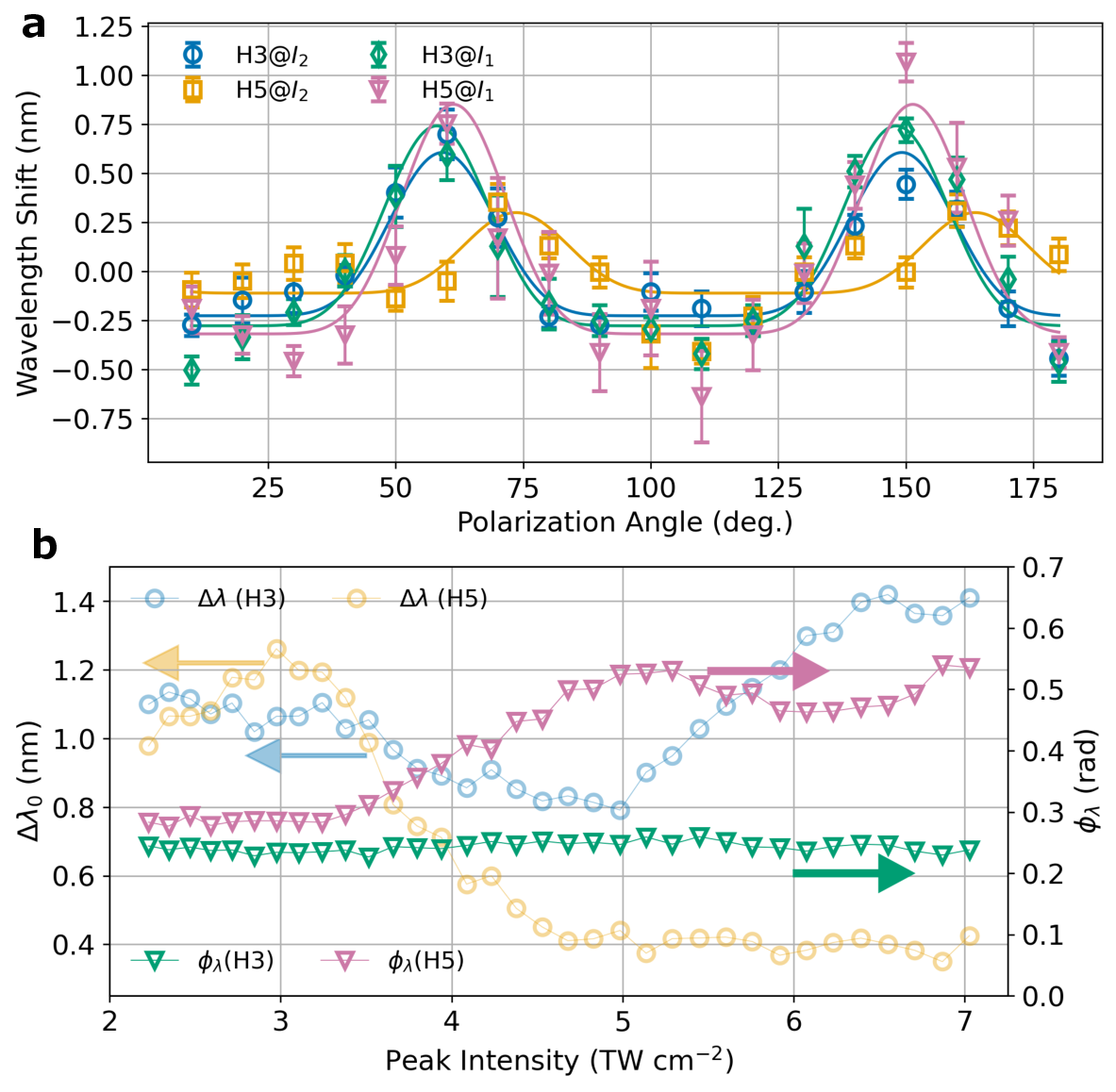}
    \caption{Analysis of orientation-dependent wavelength shifts of the third and the fifth harmonic. (a). $\Delta \lambda$ of the third and fifth harmonic together with numerical fits according to Eq.~\ref{eq:fit} as a function of the crystal orientation for two different SWIR peak intensities ($I_1 =\,$\SI{2.8}{\tera\watt\per\centi\meter\squared} and $I_2 =\,$\SI{4.7}{\tera\watt\per\centi\meter\squared}). (b) Extracted amplitudes and phases of the third and fifth harmonic wavelength shifts as a function of peak intensity.}
    \label{fig:fig_5}
\end{figure}

To further analyze the observed transition in angular dependence of the central emission wavelength shown in Fig.~\ref{fig:fig_4}(c,d), Fig.~\ref{fig:fig_5}(a) compares the wavelength shifts of the third and fifth harmonics as a function of $\theta$ at two different SWIR peak intensities ($I_1 =\,$\SI{2.8}{\tera\watt\per\centi\meter\squared} and $I_2 =\,$\SI{4.7}{\tera\watt\per\centi\meter\squared}). While the anisotropic response of the third harmonic remains unchanged, the fifth harmonic exhibits a marked evolution: its modulation amplitude decreases, and the positions of its maxima shift by approximately \SI{20}{\degree}. 

The angular dependence $\Delta \lambda(\theta)$ was fitted at each SWIR intensity using
\begin{equation}
    \label{eq:fit}
    \Delta \lambda (\theta) = \Delta\lambda_0 \times \sin^8[2 (\theta - \phi_{\lambda})]
\end{equation}
to extract the amplitude $\Delta \lambda_0$ and phase $\phi_{\lambda}$ of the oscillatory spectral shift. The fit results, presented in Fig.~\ref{fig:fig_5}(b), confirm a constant phase for the third harmonic, consistent with the fixed peak positions in Fig.~\ref{fig:fig_5}(a). The power of $8$ in Eq.~\ref{eq:fit} is chosen to obtain the best fit to the experimental data. In contrast, the phase of the fifth harmonic remains stable up to a SWIR peak intensity of approximately \SI{3.5}{\tera\watt\per\centi\meter\squared}, beyond which it undergoes an abrupt shift accompanied by a pronounced reduction in modulation amplitude.


%
This behavior can be analyzed in the context of the dominant HHG generation mechanism. Given that the seventh harmonic is well above the ZnO bandgpap, it can be attributed to a non-perturbative, recollision-based mechanism, consistent with previous works \cite{Vampa_2014, Yue_2020}. At low intensities, third harmonic generation is primarily driven by the Kerr-type response of bound electrons. The observed transition in the fifth harmonic suggests a change in its generation mechanism - from a perturbative process (likely cascaded $\chi^{(3)}$-based four-wave-mixing \cite{Garejev_2014}) to a non-perturbative regime. Due to the different dependence of these mechanisms on the intrinsic material properties, this transition alters the orientation dependence of the emission wavelength. 

The transition from perturbative to non-perturbative HHG can also be linked to the nonlinear decrease in transmission observed in Fig.~\ref{fig:fig_2}(a). Similar nonlinear transmission behavior has been reported in prior studies \cite{Grojo_2013, Juergens_2024, Sneftrup_2024} where a gradual transmission drop signals the onset of strong-field excitation. A rapid variation of $\phi_{\lambda}(\text{H5})$ sets in at \SI{3.5}{\tera\watt\per\centi\meter\squared}, corresponding to the intensity range where strong-field excitation becomes significant, as shown in Fig.~\ref{fig:fig_2}(a). 
%


\section{Conclusion}
We have investigated the interplay between HHG and PL in bulk ZnO driven by intense SWIR and above-bandgap excitation. The HHG signal displays the expected non-perturbative intensity scaling and a pronounced blueshift with increasing excitation strength, attributed to excitation-induced refractive index changes. In contrast, the accompanying PL emission exhibits a redshift and a superlinear dependence on carrier density, consistent with phonon-assisted recombination and exciton-exciton scattering (collisional luminescence), rather than band-to-band or defect-mediated recombination. 

Our results demonstrate that the PL signal, although commonly used as a representative for excitation density in solid-state HHG experiments, is not directly proportional to the density of excited carriers. Instead, when a substantial carrier density is excited, it follows a quadratic scaling characteristic of a two-exciton recombination process until a saturation and eventual decrease is observed due to exciton-exciton annihilation. While the overall power-law dependence of the PL yield on the intensity agrees with earlier studies, our findings refine its interpretation and emphasize the role of many-body excitonic dynamics. 

A key result is the intensity-induced change in the angular dependence of the fifth harmonic's emission wavelength. At moderate intensities, its anisotropic spectral shift suggests a perturbative, symmetry-sensitive generation pathway. Beyond a critical intensity, this response abruptly transitions to an isotropic regime with suppressed modulation amplitude and shifted phase, marking a change in the underlying generation mechanism. We interpret this transition as a shift from cascaded four-wave mixing to a non-perturbative recollision-type HHG process. This provides a new spectroscopic handle to identify and distinguish competing microscopic HHG mechanisms through their spectral signatures. 
%

Our findings establish a comprehensive picture of coherent and incoherent light emission in a prototypical wide-bandgap semiconductor under strong-field excitation. They demonstrate the value of analyzing not only signal yields but also spectral and angular characteristics to access microscopic generation pathways. Further studies may leverage these observables to disentangle nonlinear processes in more complex materials and to benchmark theoretical models of strong-field dynamics. Moreover, the identification of exciton-exciton scattering as the origin of the PL signal at high carrier concentration opens opportunities for exploiting this process in ultrafast luminescence-based field metrology and for studying carrier correlations in the high-density regime.

\bibliography{references}

\begin{thebibliography}{51}%
\makeatletter
\providecommand \@ifxundefined [1]{%
 \@ifx{#1\undefined}
}%
\providecommand \@ifnum [1]{%
 \ifnum #1\expandafter \@firstoftwo
 \else \expandafter \@secondoftwo
 \fi
}%
\providecommand \@ifx [1]{%
 \ifx #1\expandafter \@firstoftwo
 \else \expandafter \@secondoftwo
 \fi
}%
\providecommand \natexlab [1]{#1}%
\providecommand \enquote  [1]{``#1''}%
\providecommand \bibnamefont  [1]{#1}%
\providecommand \bibfnamefont [1]{#1}%
\providecommand \citenamefont [1]{#1}%
\providecommand \href@noop [0]{\@secondoftwo}%
\providecommand \href [0]{\begingroup \@sanitize@url \@href}%
\providecommand \@href[1]{\@@startlink{#1}\@@href}%
\providecommand \@@href[1]{\endgroup#1\@@endlink}%
\providecommand \@sanitize@url [0]{\catcode `\\12\catcode `\$12\catcode
  `\&12\catcode `\#12\catcode `\^12\catcode `\_12\catcode `\%12\relax}%
\providecommand \@@startlink[1]{}%
\providecommand \@@endlink[0]{}%
\providecommand \url  [0]{\begingroup\@sanitize@url \@url }%
\providecommand \@url [1]{\endgroup\@href {#1}{\urlprefix }}%
\providecommand \urlprefix  [0]{URL }%
\providecommand \Eprint [0]{\href }%
\providecommand \doibase [0]{https://doi.org/}%
\providecommand \selectlanguage [0]{\@gobble}%
\providecommand \bibinfo  [0]{\@secondoftwo}%
\providecommand \bibfield  [0]{\@secondoftwo}%
\providecommand \translation [1]{[#1]}%
\providecommand \BibitemOpen [0]{}%
\providecommand \bibitemStop [0]{}%
\providecommand \bibitemNoStop [0]{.\EOS\space}%
\providecommand \EOS [0]{\spacefactor3000\relax}%
\providecommand \BibitemShut  [1]{\csname bibitem#1\endcsname}%
\let\auto@bib@innerbib\@empty
\bibitem [{\citenamefont {Chin}\ \emph {et~al.}(2001)\citenamefont {Chin},
  \citenamefont {Calder{\'o}n},\ and\ \citenamefont {Kono}}]{Chin_2001}%
  \BibitemOpen
  \bibfield  {author} {\bibinfo {author} {\bibfnamefont {A.~H.}\ \bibnamefont
  {Chin}}, \bibinfo {author} {\bibfnamefont {O.~G.}\ \bibnamefont
  {Calder{\'o}n}},\ and\ \bibinfo {author} {\bibfnamefont {J.}~\bibnamefont
  {Kono}},\ }\bibfield  {title} {\bibinfo {title} {Extreme midinfrared
  nonlinear optics in semiconductors},\ }\href
  {https://doi.org/10.1103/PhysRevLett.86.3292} {\bibfield  {journal} {\bibinfo
   {journal} {Physical Review Letters}\ }\textbf {\bibinfo {volume} {86}},\
  \bibinfo {pages} {3292} (\bibinfo {year} {2001})}\BibitemShut {NoStop}%
\bibitem [{\citenamefont {Ghimire}\ \emph {et~al.}(2011)\citenamefont
  {Ghimire}, \citenamefont {DiChiara}, \citenamefont {Sistrunk}, \citenamefont
  {Agostini}, \citenamefont {DiMauro},\ and\ \citenamefont
  {Reis}}]{Ghimire_2011}%
  \BibitemOpen
  \bibfield  {author} {\bibinfo {author} {\bibfnamefont {S.}~\bibnamefont
  {Ghimire}}, \bibinfo {author} {\bibfnamefont {A.~D.}\ \bibnamefont
  {DiChiara}}, \bibinfo {author} {\bibfnamefont {E.}~\bibnamefont {Sistrunk}},
  \bibinfo {author} {\bibfnamefont {P.}~\bibnamefont {Agostini}}, \bibinfo
  {author} {\bibfnamefont {L.~F.}\ \bibnamefont {DiMauro}},\ and\ \bibinfo
  {author} {\bibfnamefont {D.~A.}\ \bibnamefont {Reis}},\ }\bibfield  {title}
  {\bibinfo {title} {Observation of high-order harmonic generation in a bulk
  crystal},\ }\href {https://doi.org/10.1038/nphys1847} {\bibfield  {journal}
  {\bibinfo  {journal} {Nature Physics}\ }\textbf {\bibinfo {volume} {7}},\
  \bibinfo {pages} {138} (\bibinfo {year} {2011})}\BibitemShut {NoStop}%
\bibitem [{\citenamefont {Li}\ \emph {et~al.}(2020)\citenamefont {Li},
  \citenamefont {Lu}, \citenamefont {Chew}, \citenamefont {Han}, \citenamefont
  {Li}, \citenamefont {Wu}, \citenamefont {Wang}, \citenamefont {Ghimire},\
  and\ \citenamefont {Chang}}]{Li_2020}%
  \BibitemOpen
  \bibfield  {author} {\bibinfo {author} {\bibfnamefont {J.}~\bibnamefont
  {Li}}, \bibinfo {author} {\bibfnamefont {J.}~\bibnamefont {Lu}}, \bibinfo
  {author} {\bibfnamefont {A.}~\bibnamefont {Chew}}, \bibinfo {author}
  {\bibfnamefont {S.}~\bibnamefont {Han}}, \bibinfo {author} {\bibfnamefont
  {J.}~\bibnamefont {Li}}, \bibinfo {author} {\bibfnamefont {Y.}~\bibnamefont
  {Wu}}, \bibinfo {author} {\bibfnamefont {H.}~\bibnamefont {Wang}}, \bibinfo
  {author} {\bibfnamefont {S.}~\bibnamefont {Ghimire}},\ and\ \bibinfo {author}
  {\bibfnamefont {Z.}~\bibnamefont {Chang}},\ }\bibfield  {title} {\bibinfo
  {title} {Attosecond science based on high harmonic generation from gases and
  solids},\ }\href {https://doi.org/10.1038/s41467-020-16480-6} {\bibfield
  {journal} {\bibinfo  {journal} {Nature Communications}\ }\textbf {\bibinfo
  {volume} {11}},\ \bibinfo {pages} {2748} (\bibinfo {year}
  {2020})}\BibitemShut {NoStop}%
\bibitem [{\citenamefont {Ossiander}\ \emph {et~al.}(2022)\citenamefont
  {Ossiander}, \citenamefont {Golyari}, \citenamefont {Scharl}, \citenamefont
  {Lehnert}, \citenamefont {Siegrist}, \citenamefont {B{\"u}rger},
  \citenamefont {Zimin}, \citenamefont {Gessner}, \citenamefont {Weidman},
  \citenamefont {Floss} \emph {et~al.}}]{Ossiander_2022}%
  \BibitemOpen
  \bibfield  {author} {\bibinfo {author} {\bibfnamefont {M.}~\bibnamefont
  {Ossiander}}, \bibinfo {author} {\bibfnamefont {K.}~\bibnamefont {Golyari}},
  \bibinfo {author} {\bibfnamefont {K.}~\bibnamefont {Scharl}}, \bibinfo
  {author} {\bibfnamefont {L.}~\bibnamefont {Lehnert}}, \bibinfo {author}
  {\bibfnamefont {F.}~\bibnamefont {Siegrist}}, \bibinfo {author}
  {\bibfnamefont {J.}~\bibnamefont {B{\"u}rger}}, \bibinfo {author}
  {\bibfnamefont {D.}~\bibnamefont {Zimin}}, \bibinfo {author} {\bibfnamefont
  {J.}~\bibnamefont {Gessner}}, \bibinfo {author} {\bibfnamefont
  {M.}~\bibnamefont {Weidman}}, \bibinfo {author} {\bibfnamefont
  {I.}~\bibnamefont {Floss}}, \emph {et~al.},\ }\bibfield  {title} {\bibinfo
  {title} {The speed limit of optoelectronics},\ }\href
  {https://doi.org/10.1038/s41467-022-29252-1} {\bibfield  {journal} {\bibinfo
  {journal} {Nature Communications}\ }\textbf {\bibinfo {volume} {13}},\
  \bibinfo {pages} {1620} (\bibinfo {year} {2022})}\BibitemShut {NoStop}%
\bibitem [{\citenamefont {Heide}\ \emph {et~al.}(2024)\citenamefont {Heide},
  \citenamefont {Keathley},\ and\ \citenamefont {Kling}}]{Heide_2024}%
  \BibitemOpen
  \bibfield  {author} {\bibinfo {author} {\bibfnamefont {C.}~\bibnamefont
  {Heide}}, \bibinfo {author} {\bibfnamefont {P.~D.}\ \bibnamefont
  {Keathley}},\ and\ \bibinfo {author} {\bibfnamefont {M.~F.}\ \bibnamefont
  {Kling}},\ }\bibfield  {title} {\bibinfo {title} {Petahertz electronics},\
  }\href {https://doi.org/10.1038/s42254-024-00764-7} {\bibfield  {journal}
  {\bibinfo  {journal} {Nature Reviews Physics}\ }\textbf {\bibinfo {volume}
  {6}},\ \bibinfo {pages} {648} (\bibinfo {year} {2024})}\BibitemShut {NoStop}%
\bibitem [{\citenamefont {Zimin}\ \emph {et~al.}(2021)\citenamefont {Zimin},
  \citenamefont {Weidman}, \citenamefont {Sch{\"o}tz}, \citenamefont {Kling},
  \citenamefont {Yakovlev}, \citenamefont {Krausz},\ and\ \citenamefont
  {Karpowicz}}]{Zimin_2021}%
  \BibitemOpen
  \bibfield  {author} {\bibinfo {author} {\bibfnamefont {D.}~\bibnamefont
  {Zimin}}, \bibinfo {author} {\bibfnamefont {M.}~\bibnamefont {Weidman}},
  \bibinfo {author} {\bibfnamefont {J.}~\bibnamefont {Sch{\"o}tz}}, \bibinfo
  {author} {\bibfnamefont {M.~F.}\ \bibnamefont {Kling}}, \bibinfo {author}
  {\bibfnamefont {V.~S.}\ \bibnamefont {Yakovlev}}, \bibinfo {author}
  {\bibfnamefont {F.}~\bibnamefont {Krausz}},\ and\ \bibinfo {author}
  {\bibfnamefont {N.}~\bibnamefont {Karpowicz}},\ }\bibfield  {title} {\bibinfo
  {title} {Petahertz-scale nonlinear photoconductive sampling in air},\ }\href
  {https://doi.org/10.1364/OPTICA.411434} {\bibfield  {journal} {\bibinfo
  {journal} {Optica}\ }\textbf {\bibinfo {volume} {8}},\ \bibinfo {pages} {586}
  (\bibinfo {year} {2021})}\BibitemShut {NoStop}%
\bibitem [{\citenamefont {Sederberg}\ \emph {et~al.}(2020)\citenamefont
  {Sederberg}, \citenamefont {Zimin}, \citenamefont {Keiber}, \citenamefont
  {Siegrist}, \citenamefont {Wismer}, \citenamefont {Yakovlev}, \citenamefont
  {Floss}, \citenamefont {Lemell}, \citenamefont {Burgd{\"o}rfer},
  \citenamefont {Schultze} \emph {et~al.}}]{Sederberg_2020}%
  \BibitemOpen
  \bibfield  {author} {\bibinfo {author} {\bibfnamefont {S.}~\bibnamefont
  {Sederberg}}, \bibinfo {author} {\bibfnamefont {D.}~\bibnamefont {Zimin}},
  \bibinfo {author} {\bibfnamefont {S.}~\bibnamefont {Keiber}}, \bibinfo
  {author} {\bibfnamefont {F.}~\bibnamefont {Siegrist}}, \bibinfo {author}
  {\bibfnamefont {M.~S.}\ \bibnamefont {Wismer}}, \bibinfo {author}
  {\bibfnamefont {V.~S.}\ \bibnamefont {Yakovlev}}, \bibinfo {author}
  {\bibfnamefont {I.}~\bibnamefont {Floss}}, \bibinfo {author} {\bibfnamefont
  {C.}~\bibnamefont {Lemell}}, \bibinfo {author} {\bibfnamefont
  {J.}~\bibnamefont {Burgd{\"o}rfer}}, \bibinfo {author} {\bibfnamefont
  {M.}~\bibnamefont {Schultze}}, \emph {et~al.},\ }\bibfield  {title} {\bibinfo
  {title} {Attosecond optoelectronic field measurement in solids},\ }\href
  {https://doi.org/10.1038/s41467-019-14268-x} {\bibfield  {journal} {\bibinfo
  {journal} {Nature Communications}\ }\textbf {\bibinfo {volume} {11}},\
  \bibinfo {pages} {430} (\bibinfo {year} {2020})}\BibitemShut {NoStop}%
\bibitem [{\citenamefont {Luu}\ \emph {et~al.}(2015)\citenamefont {Luu},
  \citenamefont {Garg}, \citenamefont {Kruchinin}, \citenamefont {Moulet},
  \citenamefont {Hassan},\ and\ \citenamefont {Goulielmakis}}]{Luu_2015}%
  \BibitemOpen
  \bibfield  {author} {\bibinfo {author} {\bibfnamefont {T.~T.}\ \bibnamefont
  {Luu}}, \bibinfo {author} {\bibfnamefont {M.}~\bibnamefont {Garg}}, \bibinfo
  {author} {\bibfnamefont {S.~Y.}\ \bibnamefont {Kruchinin}}, \bibinfo {author}
  {\bibfnamefont {A.}~\bibnamefont {Moulet}}, \bibinfo {author} {\bibfnamefont
  {M.~T.}\ \bibnamefont {Hassan}},\ and\ \bibinfo {author} {\bibfnamefont
  {E.}~\bibnamefont {Goulielmakis}},\ }\bibfield  {title} {\bibinfo {title}
  {Extreme ultraviolet high-harmonic spectroscopy of solids},\ }\href
  {https://doi.org/10.1038/nature14456} {\bibfield  {journal} {\bibinfo
  {journal} {Nature}\ }\textbf {\bibinfo {volume} {521}},\ \bibinfo {pages}
  {498} (\bibinfo {year} {2015})}\BibitemShut {NoStop}%
\bibitem [{\citenamefont {Lanin}\ \emph {et~al.}(2017)\citenamefont {Lanin},
  \citenamefont {Stepanov}, \citenamefont {Fedotov},\ and\ \citenamefont
  {Zheltikov}}]{Lanin_2017}%
  \BibitemOpen
  \bibfield  {author} {\bibinfo {author} {\bibfnamefont {A.}~\bibnamefont
  {Lanin}}, \bibinfo {author} {\bibfnamefont {E.}~\bibnamefont {Stepanov}},
  \bibinfo {author} {\bibfnamefont {A.}~\bibnamefont {Fedotov}},\ and\ \bibinfo
  {author} {\bibfnamefont {A.}~\bibnamefont {Zheltikov}},\ }\bibfield  {title}
  {\bibinfo {title} {Mapping the electron band structure by intraband
  high-harmonic generation in solids},\ }\href
  {https://doi.org/10.1364/OPTICA.4.000516} {\bibfield  {journal} {\bibinfo
  {journal} {Optica}\ }\textbf {\bibinfo {volume} {4}},\ \bibinfo {pages} {516}
  (\bibinfo {year} {2017})}\BibitemShut {NoStop}%
\bibitem [{\citenamefont {Zaks}\ \emph {et~al.}(2012)\citenamefont {Zaks},
  \citenamefont {Liu},\ and\ \citenamefont {Sherwin}}]{Zaks_2012}%
  \BibitemOpen
  \bibfield  {author} {\bibinfo {author} {\bibfnamefont {B.}~\bibnamefont
  {Zaks}}, \bibinfo {author} {\bibfnamefont {R.-B.}\ \bibnamefont {Liu}},\ and\
  \bibinfo {author} {\bibfnamefont {M.~S.}\ \bibnamefont {Sherwin}},\
  }\bibfield  {title} {\bibinfo {title} {Experimental observation of
  electron--hole recollisions},\ }\href {https://doi.org/10.1038/nature10864}
  {\bibfield  {journal} {\bibinfo  {journal} {Nature}\ }\textbf {\bibinfo
  {volume} {483}},\ \bibinfo {pages} {580} (\bibinfo {year}
  {2012})}\BibitemShut {NoStop}%
\bibitem [{\citenamefont {Uchida}\ and\ \citenamefont
  {Tanaka}(2024)}]{Uchida_2024}%
  \BibitemOpen
  \bibfield  {author} {\bibinfo {author} {\bibfnamefont {K.}~\bibnamefont
  {Uchida}}\ and\ \bibinfo {author} {\bibfnamefont {K.}~\bibnamefont
  {Tanaka}},\ }\bibfield  {title} {\bibinfo {title} {{High harmonic
  Mach--Zehnder interferometer for probing sub-laser-cycle electron dynamics in
  solids}},\ }\href {https://doi.org/10.1364/OPTICA.527675} {\bibfield
  {journal} {\bibinfo  {journal} {Optica}\ }\textbf {\bibinfo {volume} {11}},\
  \bibinfo {pages} {1130} (\bibinfo {year} {2024})}\BibitemShut {NoStop}%
\bibitem [{\citenamefont {Heide}\ \emph {et~al.}(2022)\citenamefont {Heide},
  \citenamefont {Kobayashi}, \citenamefont {Johnson}, \citenamefont {Liu},
  \citenamefont {Heinz}, \citenamefont {Reis},\ and\ \citenamefont
  {Ghimire}}]{Heide_2022}%
  \BibitemOpen
  \bibfield  {author} {\bibinfo {author} {\bibfnamefont {C.}~\bibnamefont
  {Heide}}, \bibinfo {author} {\bibfnamefont {Y.}~\bibnamefont {Kobayashi}},
  \bibinfo {author} {\bibfnamefont {A.~C.}\ \bibnamefont {Johnson}}, \bibinfo
  {author} {\bibfnamefont {F.}~\bibnamefont {Liu}}, \bibinfo {author}
  {\bibfnamefont {T.~F.}\ \bibnamefont {Heinz}}, \bibinfo {author}
  {\bibfnamefont {D.~A.}\ \bibnamefont {Reis}},\ and\ \bibinfo {author}
  {\bibfnamefont {S.}~\bibnamefont {Ghimire}},\ }\bibfield  {title} {\bibinfo
  {title} {{Probing electron-hole coherence in strongly driven 2D materials
  using high-harmonic generation}},\ }\href
  {https://doi.org/10.1364/OPTICA.444105} {\bibfield  {journal} {\bibinfo
  {journal} {Optica}\ }\textbf {\bibinfo {volume} {9}},\ \bibinfo {pages} {512}
  (\bibinfo {year} {2022})}\BibitemShut {NoStop}%
\bibitem [{\citenamefont {Gholam-Mirzaei}\ \emph {et~al.}(2018)\citenamefont
  {Gholam-Mirzaei}, \citenamefont {Beetar}, \citenamefont {Chac{\'o}n},\ and\
  \citenamefont {Chini}}]{Gholam_2018}%
  \BibitemOpen
  \bibfield  {author} {\bibinfo {author} {\bibfnamefont {S.}~\bibnamefont
  {Gholam-Mirzaei}}, \bibinfo {author} {\bibfnamefont {J.~E.}\ \bibnamefont
  {Beetar}}, \bibinfo {author} {\bibfnamefont {A.}~\bibnamefont {Chac{\'o}n}},\
  and\ \bibinfo {author} {\bibfnamefont {M.}~\bibnamefont {Chini}},\ }\bibfield
   {title} {\bibinfo {title} {{High-harmonic generation in ZnO driven by
  self-compressed mid-infrared pulses}},\ }\href
  {https://doi.org/10.1364/JOSAB.35.000A27} {\bibfield  {journal} {\bibinfo
  {journal} {JOSA B}\ }\textbf {\bibinfo {volume} {35}},\ \bibinfo {pages}
  {A27} (\bibinfo {year} {2018})}\BibitemShut {NoStop}%
\bibitem [{\citenamefont {Hollinger}\ \emph {et~al.}(2020)\citenamefont
  {Hollinger}, \citenamefont {Herrmann}, \citenamefont {Korolev}, \citenamefont
  {Zapf}, \citenamefont {Shumakova}, \citenamefont {R{\"o}der}, \citenamefont
  {Uschmann}, \citenamefont {Pug{\v{z}}lys}, \citenamefont {Baltu{\v{s}}ka},
  \citenamefont {Z{\"u}rch} \emph {et~al.}}]{Hollinger_2020}%
  \BibitemOpen
  \bibfield  {author} {\bibinfo {author} {\bibfnamefont {R.}~\bibnamefont
  {Hollinger}}, \bibinfo {author} {\bibfnamefont {P.}~\bibnamefont {Herrmann}},
  \bibinfo {author} {\bibfnamefont {V.}~\bibnamefont {Korolev}}, \bibinfo
  {author} {\bibfnamefont {M.}~\bibnamefont {Zapf}}, \bibinfo {author}
  {\bibfnamefont {V.}~\bibnamefont {Shumakova}}, \bibinfo {author}
  {\bibfnamefont {R.}~\bibnamefont {R{\"o}der}}, \bibinfo {author}
  {\bibfnamefont {I.}~\bibnamefont {Uschmann}}, \bibinfo {author}
  {\bibfnamefont {A.}~\bibnamefont {Pug{\v{z}}lys}}, \bibinfo {author}
  {\bibfnamefont {A.}~\bibnamefont {Baltu{\v{s}}ka}}, \bibinfo {author}
  {\bibfnamefont {M.}~\bibnamefont {Z{\"u}rch}}, \emph {et~al.},\ }\bibfield
  {title} {\bibinfo {title} {{Polarization dependent excitation and high
  harmonic generation from intense mid-IR laser pulses in ZnO}},\ }\href
  {https://doi.org/10.3390/nano11010004} {\bibfield  {journal} {\bibinfo
  {journal} {Nanomaterials}\ }\textbf {\bibinfo {volume} {11}},\ \bibinfo
  {pages} {4} (\bibinfo {year} {2020})}\BibitemShut {NoStop}%
\bibitem [{\citenamefont {Jiang}\ \emph {et~al.}(2019)\citenamefont {Jiang},
  \citenamefont {Gholam-Mirzaei}, \citenamefont {Crites}, \citenamefont
  {Beetar}, \citenamefont {Singh}, \citenamefont {Lu}, \citenamefont {Chini},\
  and\ \citenamefont {Lin}}]{Jiang_2019}%
  \BibitemOpen
  \bibfield  {author} {\bibinfo {author} {\bibfnamefont {S.}~\bibnamefont
  {Jiang}}, \bibinfo {author} {\bibfnamefont {S.}~\bibnamefont
  {Gholam-Mirzaei}}, \bibinfo {author} {\bibfnamefont {E.}~\bibnamefont
  {Crites}}, \bibinfo {author} {\bibfnamefont {J.~E.}\ \bibnamefont {Beetar}},
  \bibinfo {author} {\bibfnamefont {M.}~\bibnamefont {Singh}}, \bibinfo
  {author} {\bibfnamefont {R.}~\bibnamefont {Lu}}, \bibinfo {author}
  {\bibfnamefont {M.}~\bibnamefont {Chini}},\ and\ \bibinfo {author}
  {\bibfnamefont {C.}~\bibnamefont {Lin}},\ }\bibfield  {title} {\bibinfo
  {title} {{Crystal symmetry and polarization of high-order harmonics in
  ZnO}},\ }\href {https://doi.org/10.1088/1361-6455/ab470d} {\bibfield
  {journal} {\bibinfo  {journal} {Journal of Physics B: Atomic, Molecular and
  Optical Physics}\ }\textbf {\bibinfo {volume} {52}},\ \bibinfo {pages}
  {225601} (\bibinfo {year} {2019})}\BibitemShut {NoStop}%
\bibitem [{\citenamefont {Gholam-Mirzaei}\ \emph {et~al.}(2017)\citenamefont
  {Gholam-Mirzaei}, \citenamefont {Beetar},\ and\ \citenamefont
  {Chini}}]{Gholam_2017}%
  \BibitemOpen
  \bibfield  {author} {\bibinfo {author} {\bibfnamefont {S.}~\bibnamefont
  {Gholam-Mirzaei}}, \bibinfo {author} {\bibfnamefont {J.}~\bibnamefont
  {Beetar}},\ and\ \bibinfo {author} {\bibfnamefont {M.}~\bibnamefont
  {Chini}},\ }\bibfield  {title} {\bibinfo {title} {{High harmonic generation
  in ZnO with a high-power mid-IR OPA}},\ }\bibfield  {journal} {\bibinfo
  {journal} {Applied Physics Letters}\ }\textbf {\bibinfo {volume} {110}},\
  \href {https://doi.org/10.1063/1.4975362} {10.1063/1.4975362} (\bibinfo
  {year} {2017})\BibitemShut {NoStop}%
\bibitem [{\citenamefont {{Li, Wenkai and Liu, Zhe and Shao, Beijie and Qian,
  Junyu and Li, Yanyan and Peng, Yujie and Leng, Yuxin}}(2022)}]{Li_2022}%
  \BibitemOpen
  \bibfield  {author} {\bibinfo {author} {\bibnamefont {{Li, Wenkai and Liu,
  Zhe and Shao, Beijie and Qian, Junyu and Li, Yanyan and Peng, Yujie and Leng,
  Yuxin}}},\ }\bibfield  {title} {\bibinfo {title} {{Angle-resolved high-order
  harmonics in wurtzite-type ZnO}},\ }\bibfield  {journal} {\bibinfo  {journal}
  {Journal of Applied Physics}\ }\textbf {\bibinfo {volume} {132}},\ \href
  {https://doi.org/10.1063/5.0098582} {10.1063/5.0098582} (\bibinfo {year}
  {2022})\BibitemShut {NoStop}%
\bibitem [{\citenamefont {Wang}\ \emph {et~al.}(2017)\citenamefont {Wang},
  \citenamefont {Park}, \citenamefont {Lai}, \citenamefont {Xu}, \citenamefont
  {Blaga}, \citenamefont {Yang}, \citenamefont {Agostini},\ and\ \citenamefont
  {DiMauro}}]{Wang_2017}%
  \BibitemOpen
  \bibfield  {author} {\bibinfo {author} {\bibfnamefont {Z.}~\bibnamefont
  {Wang}}, \bibinfo {author} {\bibfnamefont {H.}~\bibnamefont {Park}}, \bibinfo
  {author} {\bibfnamefont {Y.~H.}\ \bibnamefont {Lai}}, \bibinfo {author}
  {\bibfnamefont {J.}~\bibnamefont {Xu}}, \bibinfo {author} {\bibfnamefont
  {C.~I.}\ \bibnamefont {Blaga}}, \bibinfo {author} {\bibfnamefont
  {F.}~\bibnamefont {Yang}}, \bibinfo {author} {\bibfnamefont {P.}~\bibnamefont
  {Agostini}},\ and\ \bibinfo {author} {\bibfnamefont {L.~F.}\ \bibnamefont
  {DiMauro}},\ }\bibfield  {title} {\bibinfo {title} {The roles of
  photo-carrier doping and driving wavelength in high harmonic generation from
  a semiconductor},\ }\href {https://doi.org/10.1038/s41467-017-01899-1}
  {\bibfield  {journal} {\bibinfo  {journal} {Nature Communications}\ }\textbf
  {\bibinfo {volume} {8}},\ \bibinfo {pages} {1} (\bibinfo {year}
  {2017})}\BibitemShut {NoStop}%
\bibitem [{\citenamefont {Xu}\ \emph {et~al.}(2022)\citenamefont {Xu},
  \citenamefont {Zhang}, \citenamefont {Yu}, \citenamefont {Han}, \citenamefont
  {Wang},\ and\ \citenamefont {Hu}}]{Xu_2022}%
  \BibitemOpen
  \bibfield  {author} {\bibinfo {author} {\bibfnamefont {S.}~\bibnamefont
  {Xu}}, \bibinfo {author} {\bibfnamefont {H.}~\bibnamefont {Zhang}}, \bibinfo
  {author} {\bibfnamefont {J.}~\bibnamefont {Yu}}, \bibinfo {author}
  {\bibfnamefont {Y.}~\bibnamefont {Han}}, \bibinfo {author} {\bibfnamefont
  {Z.}~\bibnamefont {Wang}},\ and\ \bibinfo {author} {\bibfnamefont
  {J.}~\bibnamefont {Hu}},\ }\bibfield  {title} {\bibinfo {title} {{Ultrafast
  modulation of a high harmonic generation in a bulk ZnO single crystal}},\
  }\href {https://doi.org/10.1364/OE.462638} {\bibfield  {journal} {\bibinfo
  {journal} {Optics Express}\ }\textbf {\bibinfo {volume} {30}},\ \bibinfo
  {pages} {41350} (\bibinfo {year} {2022})}\BibitemShut {NoStop}%
\bibitem [{\citenamefont {Nie}\ \emph {et~al.}(2024)\citenamefont {Nie},
  \citenamefont {Murzyn}, \citenamefont {Guery}, \citenamefont {van~den
  Hooven},\ and\ \citenamefont {Kraus}}]{Nie_2024}%
  \BibitemOpen
  \bibfield  {author} {\bibinfo {author} {\bibfnamefont {Z.}~\bibnamefont
  {Nie}}, \bibinfo {author} {\bibfnamefont {K.}~\bibnamefont {Murzyn}},
  \bibinfo {author} {\bibfnamefont {L.}~\bibnamefont {Guery}}, \bibinfo
  {author} {\bibfnamefont {T.~J.}\ \bibnamefont {van~den Hooven}},\ and\
  \bibinfo {author} {\bibfnamefont {P.~M.}\ \bibnamefont {Kraus}},\ }\bibfield
  {title} {\bibinfo {title} {{Ultrafast Permittivity Engineering Enables
  Broadband Enhancement and Spatial Emission Control of Harmonic Generation in
  ZnO}},\ }\href {https://doi.org/10.1021/acsphotonics.4c01737} {\bibfield
  {journal} {\bibinfo  {journal} {ACS Photonics}\ }\textbf {\bibinfo {volume}
  {11}},\ \bibinfo {pages} {5084} (\bibinfo {year} {2024})}\BibitemShut
  {NoStop}%
\bibitem [{\citenamefont {Liu}\ \emph {et~al.}(2021)\citenamefont {Liu},
  \citenamefont {Gholam-Mirzaei}, \citenamefont {Beetar}, \citenamefont
  {Nesper}, \citenamefont {Yousif}, \citenamefont {Nrisimhamurty},\ and\
  \citenamefont {Chini}}]{Liu_2021}%
  \BibitemOpen
  \bibfield  {author} {\bibinfo {author} {\bibfnamefont {Y.}~\bibnamefont
  {Liu}}, \bibinfo {author} {\bibfnamefont {S.}~\bibnamefont {Gholam-Mirzaei}},
  \bibinfo {author} {\bibfnamefont {J.~E.}\ \bibnamefont {Beetar}}, \bibinfo
  {author} {\bibfnamefont {J.}~\bibnamefont {Nesper}}, \bibinfo {author}
  {\bibfnamefont {A.}~\bibnamefont {Yousif}}, \bibinfo {author} {\bibfnamefont
  {M.}~\bibnamefont {Nrisimhamurty}},\ and\ \bibinfo {author} {\bibfnamefont
  {M.}~\bibnamefont {Chini}},\ }\bibfield  {title} {\bibinfo {title}
  {All-optical sampling of few-cycle infrared pulses using tunneling in a
  solid},\ }\href {https://doi.org/10.1364/PRJ.420916} {\bibfield  {journal}
  {\bibinfo  {journal} {Photonics Research}\ }\textbf {\bibinfo {volume} {9}},\
  \bibinfo {pages} {929} (\bibinfo {year} {2021})}\BibitemShut {NoStop}%
\bibitem [{\citenamefont {Truong}\ \emph {et~al.}(2025)\citenamefont {Truong},
  \citenamefont {Liu}, \citenamefont {Khatri}, \citenamefont {Zhang},
  \citenamefont {Shim},\ and\ \citenamefont {Chini}}]{Truong_2025}%
  \BibitemOpen
  \bibfield  {author} {\bibinfo {author} {\bibfnamefont {T.-C.}\ \bibnamefont
  {Truong}}, \bibinfo {author} {\bibfnamefont {Y.}~\bibnamefont {Liu}},
  \bibinfo {author} {\bibfnamefont {D.}~\bibnamefont {Khatri}}, \bibinfo
  {author} {\bibfnamefont {Y.}~\bibnamefont {Zhang}}, \bibinfo {author}
  {\bibfnamefont {B.}~\bibnamefont {Shim}},\ and\ \bibinfo {author}
  {\bibfnamefont {M.}~\bibnamefont {Chini}},\ }\bibfield  {title} {\bibinfo
  {title} {Scanless laser waveform measurement in the near-infrared},\
  }\bibfield  {journal} {\bibinfo  {journal} {APL Photonics}\ }\textbf
  {\bibinfo {volume} {10}},\ \href {https://doi.org/10.1063/5.0239294}
  {10.1063/5.0239294} (\bibinfo {year} {2025})\BibitemShut {NoStop}%
\bibitem [{\citenamefont {Juergens}\ \emph {et~al.}(2024)\citenamefont
  {Juergens}, \citenamefont {Roscam~Abbing}, \citenamefont {Mero},
  \citenamefont {Garcia}, \citenamefont {Brown}, \citenamefont {Vrakking},
  \citenamefont {Mermillod-Blondin}, \citenamefont {Kraus},\ and\ \citenamefont
  {Husakou}}]{Juergens_2024}%
  \BibitemOpen
  \bibfield  {author} {\bibinfo {author} {\bibfnamefont {P.}~\bibnamefont
  {Juergens}}, \bibinfo {author} {\bibfnamefont {S.~D.}\ \bibnamefont
  {Roscam~Abbing}}, \bibinfo {author} {\bibfnamefont {M.}~\bibnamefont {Mero}},
  \bibinfo {author} {\bibfnamefont {C.~L.}\ \bibnamefont {Garcia}}, \bibinfo
  {author} {\bibfnamefont {G.~G.}\ \bibnamefont {Brown}}, \bibinfo {author}
  {\bibfnamefont {M.~J.}\ \bibnamefont {Vrakking}}, \bibinfo {author}
  {\bibfnamefont {A.}~\bibnamefont {Mermillod-Blondin}}, \bibinfo {author}
  {\bibfnamefont {P.~M.}\ \bibnamefont {Kraus}},\ and\ \bibinfo {author}
  {\bibfnamefont {A.}~\bibnamefont {Husakou}},\ }\bibfield  {title} {\bibinfo
  {title} {{Linking High-Harmonic Generation and Strong-Field Ionization in
  Bulk Crystals}},\ }\href {https://doi.org/10.1021/acsphotonics.3c01436}
  {\bibfield  {journal} {\bibinfo  {journal} {ACS Photonics}\ }\textbf
  {\bibinfo {volume} {11}},\ \bibinfo {pages} {247} (\bibinfo {year}
  {2024})}\BibitemShut {NoStop}%
\bibitem [{\citenamefont {Grim}\ \emph {et~al.}(2013)\citenamefont {Grim},
  \citenamefont {Ucer}, \citenamefont {Burger}, \citenamefont {Bhattacharya},
  \citenamefont {Tupitsyn}, \citenamefont {Rowe}, \citenamefont {Buliga},
  \citenamefont {Trefilova}, \citenamefont {Gektin}, \citenamefont {Bizarri},
  \citenamefont {Moses},\ and\ \citenamefont {Williams}}]{Grim_2013}%
  \BibitemOpen
  \bibfield  {author} {\bibinfo {author} {\bibfnamefont {J.~Q.}\ \bibnamefont
  {Grim}}, \bibinfo {author} {\bibfnamefont {K.~B.}\ \bibnamefont {Ucer}},
  \bibinfo {author} {\bibfnamefont {A.}~\bibnamefont {Burger}}, \bibinfo
  {author} {\bibfnamefont {P.}~\bibnamefont {Bhattacharya}}, \bibinfo {author}
  {\bibfnamefont {E.}~\bibnamefont {Tupitsyn}}, \bibinfo {author}
  {\bibfnamefont {E.}~\bibnamefont {Rowe}}, \bibinfo {author} {\bibfnamefont
  {V.~M.}\ \bibnamefont {Buliga}}, \bibinfo {author} {\bibfnamefont
  {L.}~\bibnamefont {Trefilova}}, \bibinfo {author} {\bibfnamefont
  {A.}~\bibnamefont {Gektin}}, \bibinfo {author} {\bibfnamefont {G.~A.}\
  \bibnamefont {Bizarri}}, \bibinfo {author} {\bibfnamefont {W.~W.}\
  \bibnamefont {Moses}},\ and\ \bibinfo {author} {\bibfnamefont {R.~T.}\
  \bibnamefont {Williams}},\ }\bibfield  {title} {\bibinfo {title} {Nonlinear
  quenching of densely excited states in wide-gap solids},\ }\href
  {https://doi.org/10.1103/PhysRevB.87.125117} {\bibfield  {journal} {\bibinfo
  {journal} {Phys. Rev. B}\ }\textbf {\bibinfo {volume} {87}},\ \bibinfo
  {pages} {125117} (\bibinfo {year} {2013})}\BibitemShut {NoStop}%
\bibitem [{\citenamefont {Jessen}\ \emph {et~al.}(2025)\citenamefont {Jessen},
  \citenamefont {Di~Giacomo}, \citenamefont {Moreels}, \citenamefont
  {Julsgaard},\ and\ \citenamefont {Turtos}}]{Jessen_2025}%
  \BibitemOpen
  \bibfield  {author} {\bibinfo {author} {\bibfnamefont {S.}~\bibnamefont
  {Jessen}}, \bibinfo {author} {\bibfnamefont {A.}~\bibnamefont {Di~Giacomo}},
  \bibinfo {author} {\bibfnamefont {I.}~\bibnamefont {Moreels}}, \bibinfo
  {author} {\bibfnamefont {B.}~\bibnamefont {Julsgaard}},\ and\ \bibinfo
  {author} {\bibfnamefont {R.~M.}\ \bibnamefont {Turtos}},\ }\bibfield  {title}
  {\bibinfo {title} {Nonlinear quenching of excitonic emission from
  nanoplatelet films at high excitation densities},\ }\bibfield  {journal}
  {\bibinfo  {journal} {Scientific Reports}\ }\textbf {\bibinfo {volume}
  {15}},\ \href {https://doi.org/10.1038/s41598-025-04572-6}
  {10.1038/s41598-025-04572-6} (\bibinfo {year} {2025})\BibitemShut {NoStop}%
\bibitem [{\citenamefont {Grojo}\ \emph {et~al.}(2013)\citenamefont {Grojo},
  \citenamefont {Leyder}, \citenamefont {Delaporte}, \citenamefont {Marine},
  \citenamefont {Sentis},\ and\ \citenamefont {Ut{\'e}za}}]{Grojo_2013}%
  \BibitemOpen
  \bibfield  {author} {\bibinfo {author} {\bibfnamefont {D.}~\bibnamefont
  {Grojo}}, \bibinfo {author} {\bibfnamefont {S.}~\bibnamefont {Leyder}},
  \bibinfo {author} {\bibfnamefont {P.}~\bibnamefont {Delaporte}}, \bibinfo
  {author} {\bibfnamefont {W.}~\bibnamefont {Marine}}, \bibinfo {author}
  {\bibfnamefont {M.}~\bibnamefont {Sentis}},\ and\ \bibinfo {author}
  {\bibfnamefont {O.}~\bibnamefont {Ut{\'e}za}},\ }\bibfield  {title} {\bibinfo
  {title} {Long-wavelength multiphoton ionization inside band-gap solids},\
  }\href {https://doi.org/10.1103/PhysRevB.88.195135} {\bibfield  {journal}
  {\bibinfo  {journal} {Physical Review B}\ }\textbf {\bibinfo {volume} {88}},\
  \bibinfo {pages} {195135} (\bibinfo {year} {2013})}\BibitemShut {NoStop}%
\bibitem [{\citenamefont {Sneftrup}\ \emph {et~al.}(2024)\citenamefont
  {Sneftrup}, \citenamefont {Juergens}, \citenamefont {Michele}, \citenamefont
  {Andrade}, \citenamefont {Vrakking}, \citenamefont {Balling},\ and\
  \citenamefont {Mermillod-Blondin}}]{Sneftrup_2024}%
  \BibitemOpen
  \bibfield  {author} {\bibinfo {author} {\bibfnamefont {P.~S.}\ \bibnamefont
  {Sneftrup}}, \bibinfo {author} {\bibfnamefont {P.}~\bibnamefont {Juergens}},
  \bibinfo {author} {\bibfnamefont {V.~D.}\ \bibnamefont {Michele}}, \bibinfo
  {author} {\bibfnamefont {J.~R.}\ \bibnamefont {Andrade}}, \bibinfo {author}
  {\bibfnamefont {M.~J.}\ \bibnamefont {Vrakking}}, \bibinfo {author}
  {\bibfnamefont {P.}~\bibnamefont {Balling}},\ and\ \bibinfo {author}
  {\bibfnamefont {A.}~\bibnamefont {Mermillod-Blondin}},\ }\bibfield  {title}
  {\bibinfo {title} {Probing nonlinear excitation conditions: photoluminescence
  and nonlinear absorption studies in laser-irradiated dielectrics},\ }\href
  {https://doi.org/10.1007/s00339-024-07311-2} {\bibfield  {journal} {\bibinfo
  {journal} {Applied Physics A}\ }\textbf {\bibinfo {volume} {130}},\ \bibinfo
  {pages} {175} (\bibinfo {year} {2024})}\BibitemShut {NoStop}%
\bibitem [{\citenamefont {Heiland}\ \emph {et~al.}(1959)\citenamefont
  {Heiland}, \citenamefont {Mollwo},\ and\ \citenamefont
  {Stöckmann}}]{Seitz_1959}%
  \BibitemOpen
  \bibfield  {author} {\bibinfo {author} {\bibfnamefont {G.}~\bibnamefont
  {Heiland}}, \bibinfo {author} {\bibfnamefont {E.}~\bibnamefont {Mollwo}},\
  and\ \bibinfo {author} {\bibfnamefont {F.}~\bibnamefont {Stöckmann}},\
  }\bibfield  {title} {\bibinfo {title} {Electronic processes in zinc oxide}\
  }(\bibinfo  {publisher} {Academic Press},\ \bibinfo {year} {1959})\ pp.\
  \bibinfo {pages} {191--323}\BibitemShut {NoStop}%
\bibitem [{\citenamefont {Benkrima}\ \emph {et~al.}(2023)\citenamefont
  {Benkrima}, \citenamefont {Benhamida},\ and\ \citenamefont
  {Belfennache}}]{Benkrima_2023}%
  \BibitemOpen
  \bibfield  {author} {\bibinfo {author} {\bibfnamefont {Y.}~\bibnamefont
  {Benkrima}}, \bibinfo {author} {\bibfnamefont {S.}~\bibnamefont
  {Benhamida}},\ and\ \bibinfo {author} {\bibfnamefont {D.}~\bibnamefont
  {Belfennache}},\ }\bibfield  {title} {\bibinfo {title} {Theoretical study of
  structural and optical properties of zno in wurtzite phase},\ }\href
  {https://doi.org/10.15251/djnb.2023.181.11} {\bibfield  {journal} {\bibinfo
  {journal} {Digest Journal of Nanomaterials and Biostructures}\ }\textbf
  {\bibinfo {volume} {18}},\ \bibinfo {pages} {11–19} (\bibinfo {year}
  {2023})}\BibitemShut {NoStop}%
\bibitem [{\citenamefont {Kirm}\ \emph {et~al.}(2009)\citenamefont {Kirm},
  \citenamefont {Nagirnyi}, \citenamefont {Feldbach}, \citenamefont
  {De~Grazia}, \citenamefont {Carr\'e}, \citenamefont {Merdji}, \citenamefont
  {Guizard}, \citenamefont {Geoffroy}, \citenamefont {Gaudin}, \citenamefont
  {Fedorov}, \citenamefont {Martin}, \citenamefont {Vasil'ev},\ and\
  \citenamefont {Belsky}}]{Kirm_2009}%
  \BibitemOpen
  \bibfield  {author} {\bibinfo {author} {\bibfnamefont {M.}~\bibnamefont
  {Kirm}}, \bibinfo {author} {\bibfnamefont {V.}~\bibnamefont {Nagirnyi}},
  \bibinfo {author} {\bibfnamefont {E.}~\bibnamefont {Feldbach}}, \bibinfo
  {author} {\bibfnamefont {M.}~\bibnamefont {De~Grazia}}, \bibinfo {author}
  {\bibfnamefont {B.}~\bibnamefont {Carr\'e}}, \bibinfo {author} {\bibfnamefont
  {H.}~\bibnamefont {Merdji}}, \bibinfo {author} {\bibfnamefont
  {S.}~\bibnamefont {Guizard}}, \bibinfo {author} {\bibfnamefont
  {G.}~\bibnamefont {Geoffroy}}, \bibinfo {author} {\bibfnamefont
  {J.}~\bibnamefont {Gaudin}}, \bibinfo {author} {\bibfnamefont
  {N.}~\bibnamefont {Fedorov}}, \bibinfo {author} {\bibfnamefont
  {P.}~\bibnamefont {Martin}}, \bibinfo {author} {\bibfnamefont
  {A.}~\bibnamefont {Vasil'ev}},\ and\ \bibinfo {author} {\bibfnamefont
  {A.}~\bibnamefont {Belsky}},\ }\bibfield  {title} {\bibinfo {title}
  {Exciton-exciton interactions in ${\text{cdwo}}_{4}$ irradiated by intense
  femtosecond vacuum ultraviolet pulses},\ }\href
  {https://doi.org/10.1103/PhysRevB.79.233103} {\bibfield  {journal} {\bibinfo
  {journal} {Physical Review B}\ }\textbf {\bibinfo {volume} {79}},\ \bibinfo
  {pages} {233103} (\bibinfo {year} {2009})}\BibitemShut {NoStop}%
\bibitem [{\citenamefont {Spassky}\ \emph {et~al.}(2019)\citenamefont
  {Spassky}, \citenamefont {Vasil'ev}, \citenamefont {Belsky}, \citenamefont
  {Fedorov}, \citenamefont {Martin}, \citenamefont {Markov}, \citenamefont
  {Buzanov}, \citenamefont {Kozlova},\ and\ \citenamefont
  {Shlegel}}]{Spassky_2019}%
  \BibitemOpen
  \bibfield  {author} {\bibinfo {author} {\bibfnamefont {D.}~\bibnamefont
  {Spassky}}, \bibinfo {author} {\bibfnamefont {A.}~\bibnamefont {Vasil'ev}},
  \bibinfo {author} {\bibfnamefont {A.}~\bibnamefont {Belsky}}, \bibinfo
  {author} {\bibfnamefont {N.}~\bibnamefont {Fedorov}}, \bibinfo {author}
  {\bibfnamefont {P.}~\bibnamefont {Martin}}, \bibinfo {author} {\bibfnamefont
  {S.}~\bibnamefont {Markov}}, \bibinfo {author} {\bibfnamefont
  {O.}~\bibnamefont {Buzanov}}, \bibinfo {author} {\bibfnamefont
  {N.}~\bibnamefont {Kozlova}},\ and\ \bibinfo {author} {\bibfnamefont
  {V.}~\bibnamefont {Shlegel}},\ }\bibfield  {title} {\bibinfo {title}
  {{Excitation density effects in luminescence properties of CaMoO4 and
  ZnMoO4}},\ }\href
  {https://doi.org/https://doi.org/10.1016/j.optmat.2019.02.011} {\bibfield
  {journal} {\bibinfo  {journal} {Optical Materials}\ }\textbf {\bibinfo
  {volume} {90}},\ \bibinfo {pages} {7} (\bibinfo {year} {2019})}\BibitemShut
  {NoStop}%
\bibitem [{\citenamefont {{\"O}zg{\"u}r}\ \emph {et~al.}(2005)\citenamefont
  {{\"O}zg{\"u}r}, \citenamefont {Alivov}, \citenamefont {Liu}, \citenamefont
  {Teke}, \citenamefont {Reshchikov}, \citenamefont {Do{\u{g}}an},
  \citenamefont {Avrutin}, \citenamefont {Cho},\ and\ \citenamefont
  {Morko{\c{c}}}}]{Ozgur_2005}%
  \BibitemOpen
  \bibfield  {author} {\bibinfo {author} {\bibfnamefont {{\"U}.}~\bibnamefont
  {{\"O}zg{\"u}r}}, \bibinfo {author} {\bibfnamefont {Y.~I.}\ \bibnamefont
  {Alivov}}, \bibinfo {author} {\bibfnamefont {C.}~\bibnamefont {Liu}},
  \bibinfo {author} {\bibfnamefont {A.}~\bibnamefont {Teke}}, \bibinfo {author}
  {\bibfnamefont {M.~A.}\ \bibnamefont {Reshchikov}}, \bibinfo {author}
  {\bibfnamefont {S.}~\bibnamefont {Do{\u{g}}an}}, \bibinfo {author}
  {\bibfnamefont {V.~C. S.~J.}\ \bibnamefont {Avrutin}}, \bibinfo {author}
  {\bibfnamefont {S.-J.}\ \bibnamefont {Cho}},\ and\ \bibinfo {author}
  {\bibfnamefont {H.}~\bibnamefont {Morko{\c{c}}}},\ }\bibfield  {title}
  {\bibinfo {title} {{A comprehensive review of ZnO materials and devices}},\
  }\bibfield  {journal} {\bibinfo  {journal} {Journal of Applied Physics}\
  }\textbf {\bibinfo {volume} {98}},\ \href {https://doi.org/10.1063/1.1992666}
  {10.1063/1.1992666} (\bibinfo {year} {2005})\BibitemShut {NoStop}%
\bibitem [{\citenamefont {Alivov}\ \emph {et~al.}(2003)\citenamefont {Alivov},
  \citenamefont {Kalinina}, \citenamefont {Cherenkov}, \citenamefont {Look},
  \citenamefont {Ataev}, \citenamefont {Omaev}, \citenamefont {Chukichev},\
  and\ \citenamefont {Bagnall}}]{Alivov_2003}%
  \BibitemOpen
  \bibfield  {author} {\bibinfo {author} {\bibfnamefont {Y.~I.}\ \bibnamefont
  {Alivov}}, \bibinfo {author} {\bibfnamefont {E.}~\bibnamefont {Kalinina}},
  \bibinfo {author} {\bibfnamefont {A.}~\bibnamefont {Cherenkov}}, \bibinfo
  {author} {\bibfnamefont {D.~C.}\ \bibnamefont {Look}}, \bibinfo {author}
  {\bibfnamefont {B.}~\bibnamefont {Ataev}}, \bibinfo {author} {\bibfnamefont
  {A.}~\bibnamefont {Omaev}}, \bibinfo {author} {\bibfnamefont
  {M.}~\bibnamefont {Chukichev}},\ and\ \bibinfo {author} {\bibfnamefont
  {D.}~\bibnamefont {Bagnall}},\ }\bibfield  {title} {\bibinfo {title}
  {{Fabrication and characterization of n-ZnO/p-AlGaN heterojunction
  light-emitting diodes on 6H-SiC substrates}},\ }\href
  {https://doi.org/10.1063/1.1632537} {\bibfield  {journal} {\bibinfo
  {journal} {Applied Physics Letters}\ }\textbf {\bibinfo {volume} {83}},\
  \bibinfo {pages} {4719} (\bibinfo {year} {2003})}\BibitemShut {NoStop}%
\bibitem [{\citenamefont {Klingshirn}(2012)}]{Klingshirn_2012}%
  \BibitemOpen
  \bibfield  {author} {\bibinfo {author} {\bibfnamefont {C.~F.}\ \bibnamefont
  {Klingshirn}},\ }\href {https://doi.org/10.1007/978-3-642-28362-8} {\emph
  {\bibinfo {title} {Semiconductor optics}}}\ (\bibinfo  {publisher} {Springer
  Science \& Business Media},\ \bibinfo {year} {2012})\BibitemShut {NoStop}%
\bibitem [{\citenamefont {Chen}\ \emph {et~al.}(2001)\citenamefont {Chen},
  \citenamefont {Tuan}, \citenamefont {Segawa}, \citenamefont {Ko},
  \citenamefont {Hong},\ and\ \citenamefont {Yao}}]{Chen_2001}%
  \BibitemOpen
  \bibfield  {author} {\bibinfo {author} {\bibfnamefont {Y.}~\bibnamefont
  {Chen}}, \bibinfo {author} {\bibfnamefont {N.~T.}\ \bibnamefont {Tuan}},
  \bibinfo {author} {\bibfnamefont {Y.}~\bibnamefont {Segawa}}, \bibinfo
  {author} {\bibfnamefont {H.-j.}\ \bibnamefont {Ko}}, \bibinfo {author}
  {\bibfnamefont {S.-k.}\ \bibnamefont {Hong}},\ and\ \bibinfo {author}
  {\bibfnamefont {T.}~\bibnamefont {Yao}},\ }\bibfield  {title} {\bibinfo
  {title} {Stimulated emission and optical gain in zno epilayers grown by
  plasma-assisted molecular-beam epitaxy with buffers},\ }\href
  {https://doi.org/10.1063/1.1355665} {\bibfield  {journal} {\bibinfo
  {journal} {Applied Physics Letters}\ }\textbf {\bibinfo {volume} {78}},\
  \bibinfo {pages} {1469} (\bibinfo {year} {2001})}\BibitemShut {NoStop}%
\bibitem [{\citenamefont {Baxter}\ and\ \citenamefont
  {Schmuttenmaer}(2006)}]{Baxter_2006}%
  \BibitemOpen
  \bibfield  {author} {\bibinfo {author} {\bibfnamefont {J.~B.}\ \bibnamefont
  {Baxter}}\ and\ \bibinfo {author} {\bibfnamefont {C.~A.}\ \bibnamefont
  {Schmuttenmaer}},\ }\bibfield  {title} {\bibinfo {title} {{Conductivity of
  ZnO Nanowires, Nanoparticles, and Thin Films Using Time-Resolved Terahertz
  Spectroscopy}},\ }\href {https://doi.org/10.1021/jp064399a} {\bibfield
  {journal} {\bibinfo  {journal} {The Journal of Physical Chemistry B}\
  }\textbf {\bibinfo {volume} {110}},\ \bibinfo {pages} {25229} (\bibinfo
  {year} {2006})},\ \bibinfo {note} {pMID: 17165967}\BibitemShut {NoStop}%
\bibitem [{\citenamefont {Deinert}\ \emph {et~al.}(2014)\citenamefont
  {Deinert}, \citenamefont {Wegkamp}, \citenamefont {Meyer}, \citenamefont
  {Richter}, \citenamefont {Wolf},\ and\ \citenamefont
  {St\"ahler}}]{Deinert_2014}%
  \BibitemOpen
  \bibfield  {author} {\bibinfo {author} {\bibfnamefont {J.-C.}\ \bibnamefont
  {Deinert}}, \bibinfo {author} {\bibfnamefont {D.}~\bibnamefont {Wegkamp}},
  \bibinfo {author} {\bibfnamefont {M.}~\bibnamefont {Meyer}}, \bibinfo
  {author} {\bibfnamefont {C.}~\bibnamefont {Richter}}, \bibinfo {author}
  {\bibfnamefont {M.}~\bibnamefont {Wolf}},\ and\ \bibinfo {author}
  {\bibfnamefont {J.}~\bibnamefont {St\"ahler}},\ }\bibfield  {title} {\bibinfo
  {title} {Ultrafast exciton formation at the $\mathrm{ZnO}(10\overline{1}0)$
  surface},\ }\href {https://doi.org/10.1103/PhysRevLett.113.057602} {\bibfield
   {journal} {\bibinfo  {journal} {Phys. Rev. Lett.}\ }\textbf {\bibinfo
  {volume} {113}},\ \bibinfo {pages} {057602} (\bibinfo {year}
  {2014})}\BibitemShut {NoStop}%
\bibitem [{\citenamefont {Milne}\ \emph {et~al.}(2023)\citenamefont {Milne},
  \citenamefont {Nagornova}, \citenamefont {Pope}, \citenamefont {Chen},
  \citenamefont {Rossi}, \citenamefont {Szlachetko}, \citenamefont {Gawelda},
  \citenamefont {Britz}, \citenamefont {van Driel}, \citenamefont {Sala},
  \citenamefont {Ebner}, \citenamefont {Katayama}, \citenamefont {Southworth},
  \citenamefont {Doumy}, \citenamefont {March}, \citenamefont {Lehmann},
  \citenamefont {Mucke}, \citenamefont {Iablonskyi}, \citenamefont {Kumagai},
  \citenamefont {Knopp}, \citenamefont {Motomura}, \citenamefont {Togashi},
  \citenamefont {Owada}, \citenamefont {Yabashi}, \citenamefont {Nielsen},
  \citenamefont {Pajek}, \citenamefont {Ueda}, \citenamefont {Abela},
  \citenamefont {Penfold},\ and\ \citenamefont {Chergui}}]{Milne_2023}%
  \BibitemOpen
  \bibfield  {author} {\bibinfo {author} {\bibfnamefont {C.~J.}\ \bibnamefont
  {Milne}}, \bibinfo {author} {\bibfnamefont {N.}~\bibnamefont {Nagornova}},
  \bibinfo {author} {\bibfnamefont {T.}~\bibnamefont {Pope}}, \bibinfo {author}
  {\bibfnamefont {H.-Y.}\ \bibnamefont {Chen}}, \bibinfo {author}
  {\bibfnamefont {T.}~\bibnamefont {Rossi}}, \bibinfo {author} {\bibfnamefont
  {J.}~\bibnamefont {Szlachetko}}, \bibinfo {author} {\bibfnamefont
  {W.}~\bibnamefont {Gawelda}}, \bibinfo {author} {\bibfnamefont
  {A.}~\bibnamefont {Britz}}, \bibinfo {author} {\bibfnamefont {T.~B.}\
  \bibnamefont {van Driel}}, \bibinfo {author} {\bibfnamefont {L.}~\bibnamefont
  {Sala}}, \bibinfo {author} {\bibfnamefont {S.}~\bibnamefont {Ebner}},
  \bibinfo {author} {\bibfnamefont {T.}~\bibnamefont {Katayama}}, \bibinfo
  {author} {\bibfnamefont {S.~H.}\ \bibnamefont {Southworth}}, \bibinfo
  {author} {\bibfnamefont {G.}~\bibnamefont {Doumy}}, \bibinfo {author}
  {\bibfnamefont {A.~M.}\ \bibnamefont {March}}, \bibinfo {author}
  {\bibfnamefont {C.~S.}\ \bibnamefont {Lehmann}}, \bibinfo {author}
  {\bibfnamefont {M.}~\bibnamefont {Mucke}}, \bibinfo {author} {\bibfnamefont
  {D.}~\bibnamefont {Iablonskyi}}, \bibinfo {author} {\bibfnamefont
  {Y.}~\bibnamefont {Kumagai}}, \bibinfo {author} {\bibfnamefont
  {G.}~\bibnamefont {Knopp}}, \bibinfo {author} {\bibfnamefont
  {K.}~\bibnamefont {Motomura}}, \bibinfo {author} {\bibfnamefont
  {T.}~\bibnamefont {Togashi}}, \bibinfo {author} {\bibfnamefont
  {S.}~\bibnamefont {Owada}}, \bibinfo {author} {\bibfnamefont
  {M.}~\bibnamefont {Yabashi}}, \bibinfo {author} {\bibfnamefont {M.~M.}\
  \bibnamefont {Nielsen}}, \bibinfo {author} {\bibfnamefont {M.}~\bibnamefont
  {Pajek}}, \bibinfo {author} {\bibfnamefont {K.}~\bibnamefont {Ueda}},
  \bibinfo {author} {\bibfnamefont {R.}~\bibnamefont {Abela}}, \bibinfo
  {author} {\bibfnamefont {T.~J.}\ \bibnamefont {Penfold}},\ and\ \bibinfo
  {author} {\bibfnamefont {M.}~\bibnamefont {Chergui}},\ }\bibfield  {title}
  {\bibinfo {title} {Disentangling the evolution of electrons and holes in
  photoexcited {ZnO} nanoparticles},\ }\href
  {https://doi.org/10.1063/4.0000204} {\bibfield  {journal} {\bibinfo
  {journal} {Structural Dynamics}\ }\textbf {\bibinfo {volume} {10}},\ \bibinfo
  {pages} {064501} (\bibinfo {year} {2023})}\BibitemShut {NoStop}%
\bibitem [{\citenamefont {Wei}\ \emph {et~al.}(2021)\citenamefont {Wei},
  \citenamefont {Hei}, \citenamefont {Xu}, \citenamefont {Liu}, \citenamefont
  {Guo}, \citenamefont {Weng}, \citenamefont {Tan},\ and\ \citenamefont
  {Sheng}}]{Wei_2021}%
  \BibitemOpen
  \bibfield  {author} {\bibinfo {author} {\bibfnamefont {K.}~\bibnamefont
  {Wei}}, \bibinfo {author} {\bibfnamefont {D.}~\bibnamefont {Hei}}, \bibinfo
  {author} {\bibfnamefont {Q.}~\bibnamefont {Xu}}, \bibinfo {author}
  {\bibfnamefont {J.}~\bibnamefont {Liu}}, \bibinfo {author} {\bibfnamefont
  {Q.}~\bibnamefont {Guo}}, \bibinfo {author} {\bibfnamefont {X.}~\bibnamefont
  {Weng}}, \bibinfo {author} {\bibfnamefont {X.}~\bibnamefont {Tan}},\ and\
  \bibinfo {author} {\bibfnamefont {L.}~\bibnamefont {Sheng}},\ }\bibfield
  {title} {\bibinfo {title} {Photoluminescence nonlinearity and picosecond
  transient absorption in an {LYSO:Ce} scintillator excited by a 266 nm
  ultraviolet laser},\ }\href {https://doi.org/10.1039/D1RA00347J} {\bibfield
  {journal} {\bibinfo  {journal} {RSC Adv.}\ }\textbf {\bibinfo {volume}
  {11}},\ \bibinfo {pages} {17020} (\bibinfo {year} {2021})}\BibitemShut
  {NoStop}%
\bibitem [{\citenamefont {Sergaeva}\ \emph {et~al.}(2018)\citenamefont
  {Sergaeva}, \citenamefont {Gruzdev}, \citenamefont {Austin},\ and\
  \citenamefont {Chowdhury}}]{sergaeva2018ultrafast}%
  \BibitemOpen
  \bibfield  {author} {\bibinfo {author} {\bibfnamefont {O.}~\bibnamefont
  {Sergaeva}}, \bibinfo {author} {\bibfnamefont {V.}~\bibnamefont {Gruzdev}},
  \bibinfo {author} {\bibfnamefont {D.}~\bibnamefont {Austin}},\ and\ \bibinfo
  {author} {\bibfnamefont {E.}~\bibnamefont {Chowdhury}},\ }\bibfield  {title}
  {\bibinfo {title} {{Ultrafast excitation of conduction-band electrons by
  high-intensity ultrashort laser pulses in band-gap solids: Vinogradov
  equation versus Drude model}},\ }\href
  {https://doi.org/10.1364/JOSAB.35.002895} {\bibfield  {journal} {\bibinfo
  {journal} {Journal of the Optical Society of America B}\ }\textbf {\bibinfo
  {volume} {35}},\ \bibinfo {pages} {2895} (\bibinfo {year}
  {2018})}\BibitemShut {NoStop}%
\bibitem [{\citenamefont {Verhoef}\ \emph {et~al.}(2008)\citenamefont
  {Verhoef}, \citenamefont {Mitrofanov}, \citenamefont {Zheltikov},\ and\
  \citenamefont {Baltu{\v{s}}ka}}]{verhoef_2008}%
  \BibitemOpen
  \bibfield  {author} {\bibinfo {author} {\bibfnamefont {A.~J.}\ \bibnamefont
  {Verhoef}}, \bibinfo {author} {\bibfnamefont {A.}~\bibnamefont {Mitrofanov}},
  \bibinfo {author} {\bibfnamefont {A.}~\bibnamefont {Zheltikov}},\ and\
  \bibinfo {author} {\bibfnamefont {A.}~\bibnamefont {Baltu{\v{s}}ka}},\
  }\bibfield  {title} {\bibinfo {title} {Plasma-blueshift spectral shear
  interferometry for characterization of ultimately short optical pulses},\
  }\href {https://doi.org/10.1364/OL.34.000082} {\bibfield  {journal} {\bibinfo
   {journal} {Optics Letters}\ }\textbf {\bibinfo {volume} {34}},\ \bibinfo
  {pages} {82} (\bibinfo {year} {2008})}\BibitemShut {NoStop}%
\bibitem [{\citenamefont {van~der Geest}\ \emph {et~al.}(2023)\citenamefont
  {van~der Geest}, \citenamefont {de~Boer}, \citenamefont {Murzyn},
  \citenamefont {J{\"u}rgens}, \citenamefont {Ehrler},\ and\ \citenamefont
  {Kraus}}]{van_2023}%
  \BibitemOpen
  \bibfield  {author} {\bibinfo {author} {\bibfnamefont {M.~L.}\ \bibnamefont
  {van~der Geest}}, \bibinfo {author} {\bibfnamefont {J.~J.}\ \bibnamefont
  {de~Boer}}, \bibinfo {author} {\bibfnamefont {K.}~\bibnamefont {Murzyn}},
  \bibinfo {author} {\bibfnamefont {P.}~\bibnamefont {J{\"u}rgens}}, \bibinfo
  {author} {\bibfnamefont {B.}~\bibnamefont {Ehrler}},\ and\ \bibinfo {author}
  {\bibfnamefont {P.~M.}\ \bibnamefont {Kraus}},\ }\bibfield  {title} {\bibinfo
  {title} {Transient high-harmonic spectroscopy in an inorganic--organic lead
  halide perovskite},\ }\href {https://doi.org/10.1021/acs.jpclett.3c02588}
  {\bibfield  {journal} {\bibinfo  {journal} {The Journal of Physical Chemistry
  Letters}\ }\textbf {\bibinfo {volume} {14}},\ \bibinfo {pages} {10810}
  (\bibinfo {year} {2023})}\BibitemShut {NoStop}%
\bibitem [{\citenamefont {Koll}\ \emph {et~al.}(2025)\citenamefont {Koll},
  \citenamefont {Jensen}, \citenamefont {van Essen}, \citenamefont
  {de~Keijzer}, \citenamefont {Olsson}, \citenamefont {Cottom}, \citenamefont
  {Witting}, \citenamefont {Husakou}, \citenamefont {Vrakking}, \citenamefont
  {Madsen}, \citenamefont {Kraus},\ and\ \citenamefont
  {J{\"u}rgens}}]{Koll_2025}%
  \BibitemOpen
  \bibfield  {author} {\bibinfo {author} {\bibfnamefont {L.-M.}\ \bibnamefont
  {Koll}}, \bibinfo {author} {\bibfnamefont {S.~V.~B.}\ \bibnamefont {Jensen}},
  \bibinfo {author} {\bibfnamefont {P.~J.}\ \bibnamefont {van Essen}}, \bibinfo
  {author} {\bibfnamefont {B.}~\bibnamefont {de~Keijzer}}, \bibinfo {author}
  {\bibfnamefont {E.}~\bibnamefont {Olsson}}, \bibinfo {author} {\bibfnamefont
  {J.}~\bibnamefont {Cottom}}, \bibinfo {author} {\bibfnamefont
  {T.}~\bibnamefont {Witting}}, \bibinfo {author} {\bibfnamefont
  {A.}~\bibnamefont {Husakou}}, \bibinfo {author} {\bibfnamefont {M.~J.~J.}\
  \bibnamefont {Vrakking}}, \bibinfo {author} {\bibfnamefont {L.~B.}\
  \bibnamefont {Madsen}}, \bibinfo {author} {\bibfnamefont {P.~M.}\
  \bibnamefont {Kraus}},\ and\ \bibinfo {author} {\bibfnamefont
  {P.}~\bibnamefont {J{\"u}rgens}},\ }\href {https://arxiv.org/abs/2502.20564}
  {\bibinfo {title} {Extreme ultraviolet high-harmonic interferometry of
  excitation-induced bandgap dynamics in solids}} (\bibinfo {year} {2025}),\
  \Eprint {https://arxiv.org/abs/2502.20564} {arXiv:2502.20564
  [physics.optics]} \BibitemShut {NoStop}%
\bibitem [{\citenamefont {Reynolds}\ \emph {et~al.}(2000)\citenamefont
  {Reynolds}, \citenamefont {Look},\ and\ \citenamefont
  {Jogai}}]{Reynolds_2000}%
  \BibitemOpen
  \bibfield  {author} {\bibinfo {author} {\bibfnamefont {D.}~\bibnamefont
  {Reynolds}}, \bibinfo {author} {\bibfnamefont {D.~C.}\ \bibnamefont {Look}},\
  and\ \bibinfo {author} {\bibfnamefont {B.}~\bibnamefont {Jogai}},\ }\bibfield
   {title} {\bibinfo {title} {{Combined effects of screening and band gap
  renormalization on the energy of optical transitions in ZnO and GaN}},\
  }\href {https://doi.org/10.1063/1.1320026} {\bibfield  {journal} {\bibinfo
  {journal} {Journal of Applied Physics}\ }\textbf {\bibinfo {volume} {88}},\
  \bibinfo {pages} {5760} (\bibinfo {year} {2000})}\BibitemShut {NoStop}%
\bibitem [{\citenamefont {Dai}\ \emph {et~al.}(2014)\citenamefont {Dai},
  \citenamefont {Xu}, \citenamefont {Nakamura}, \citenamefont {Wang},
  \citenamefont {Li},\ and\ \citenamefont {Lin}}]{Dai_2014}%
  \BibitemOpen
  \bibfield  {author} {\bibinfo {author} {\bibfnamefont {J.}~\bibnamefont
  {Dai}}, \bibinfo {author} {\bibfnamefont {C.}~\bibnamefont {Xu}}, \bibinfo
  {author} {\bibfnamefont {T.}~\bibnamefont {Nakamura}}, \bibinfo {author}
  {\bibfnamefont {Y.}~\bibnamefont {Wang}}, \bibinfo {author} {\bibfnamefont
  {J.}~\bibnamefont {Li}},\ and\ \bibinfo {author} {\bibfnamefont
  {Y.}~\bibnamefont {Lin}},\ }\bibfield  {title} {\bibinfo {title}
  {{Electron--hole plasma induced band gap renormalization in ZnO microlaser
  cavities}},\ }\href {https://doi.org/10.1364/OE.22.028831} {\bibfield
  {journal} {\bibinfo  {journal} {Optics Express}\ }\textbf {\bibinfo {volume}
  {22}},\ \bibinfo {pages} {28831} (\bibinfo {year} {2014})}\BibitemShut
  {NoStop}%
\bibitem [{\citenamefont {Ziaja}\ \emph {et~al.}(2015)\citenamefont {Ziaja},
  \citenamefont {Medvedev}, \citenamefont {Tkachenko}, \citenamefont
  {Maltezopoulos},\ and\ \citenamefont {Wurth}}]{Ziaja_2015}%
  \BibitemOpen
  \bibfield  {author} {\bibinfo {author} {\bibfnamefont {B.}~\bibnamefont
  {Ziaja}}, \bibinfo {author} {\bibfnamefont {N.}~\bibnamefont {Medvedev}},
  \bibinfo {author} {\bibfnamefont {V.}~\bibnamefont {Tkachenko}}, \bibinfo
  {author} {\bibfnamefont {T.}~\bibnamefont {Maltezopoulos}},\ and\ \bibinfo
  {author} {\bibfnamefont {W.}~\bibnamefont {Wurth}},\ }\bibfield  {title}
  {\bibinfo {title} {{Time-resolved observation of band-gap shrinking and
  electron-lattice thermalization within X-ray excited gallium arsenide}},\
  }\href {https://doi.org/10.1038/srep18068} {\bibfield  {journal} {\bibinfo
  {journal} {Scientific reports}\ }\textbf {\bibinfo {volume} {5}},\ \bibinfo
  {pages} {18068} (\bibinfo {year} {2015})}\BibitemShut {NoStop}%
\bibitem [{\citenamefont {Liu}\ \emph {et~al.}(2019)\citenamefont {Liu},
  \citenamefont {Ziffer}, \citenamefont {Hansen}, \citenamefont {Wang},\ and\
  \citenamefont {Zhu}}]{Liu_2019}%
  \BibitemOpen
  \bibfield  {author} {\bibinfo {author} {\bibfnamefont {F.}~\bibnamefont
  {Liu}}, \bibinfo {author} {\bibfnamefont {M.~E.}\ \bibnamefont {Ziffer}},
  \bibinfo {author} {\bibfnamefont {K.~R.}\ \bibnamefont {Hansen}}, \bibinfo
  {author} {\bibfnamefont {J.}~\bibnamefont {Wang}},\ and\ \bibinfo {author}
  {\bibfnamefont {X.}~\bibnamefont {Zhu}},\ }\bibfield  {title} {\bibinfo
  {title} {{Direct determination of band-gap renormalization in the
  photoexcited monolayer MoS2}},\ }\href
  {https://doi.org/10.1103/PhysRevLett.122.246803} {\bibfield  {journal}
  {\bibinfo  {journal} {Physical Review Letters}\ }\textbf {\bibinfo {volume}
  {122}},\ \bibinfo {pages} {246803} (\bibinfo {year} {2019})}\BibitemShut
  {NoStop}%
\bibitem [{\citenamefont {Ren}\ \emph {et~al.}(2022)\citenamefont {Ren},
  \citenamefont {Huang},\ and\ \citenamefont {Wang}}]{Ren_2022}%
  \BibitemOpen
  \bibfield  {author} {\bibinfo {author} {\bibfnamefont {Y.}~\bibnamefont
  {Ren}}, \bibinfo {author} {\bibfnamefont {Z.}~\bibnamefont {Huang}},\ and\
  \bibinfo {author} {\bibfnamefont {Y.}~\bibnamefont {Wang}},\ }\bibfield
  {title} {\bibinfo {title} {Dynamic and giant bandgap renormalization dictates
  the transient optical response in perovskite quantum dots},\ }\bibfield
  {journal} {\bibinfo  {journal} {Applied Physics Letters}\ }\textbf {\bibinfo
  {volume} {121}},\ \href {https://doi.org/10.1063/5.0131286}
  {10.1063/5.0131286} (\bibinfo {year} {2022})\BibitemShut {NoStop}%
\bibitem [{\citenamefont {Vampa}\ \emph {et~al.}(2014)\citenamefont {Vampa},
  \citenamefont {McDonald}, \citenamefont {Orlando}, \citenamefont {Klug},
  \citenamefont {Corkum},\ and\ \citenamefont {Brabec}}]{Vampa_2014}%
  \BibitemOpen
  \bibfield  {author} {\bibinfo {author} {\bibfnamefont {G.}~\bibnamefont
  {Vampa}}, \bibinfo {author} {\bibfnamefont {C.}~\bibnamefont {McDonald}},
  \bibinfo {author} {\bibfnamefont {G.}~\bibnamefont {Orlando}}, \bibinfo
  {author} {\bibfnamefont {D.}~\bibnamefont {Klug}}, \bibinfo {author}
  {\bibfnamefont {P.}~\bibnamefont {Corkum}},\ and\ \bibinfo {author}
  {\bibfnamefont {T.}~\bibnamefont {Brabec}},\ }\bibfield  {title} {\bibinfo
  {title} {Theoretical analysis of high-harmonic generation in solids},\ }\href
  {https://doi.org/10.1103/PhysRevLett.113.073901} {\bibfield  {journal}
  {\bibinfo  {journal} {Physical Review Letters}\ }\textbf {\bibinfo {volume}
  {113}},\ \bibinfo {pages} {073901} (\bibinfo {year} {2014})}\BibitemShut
  {NoStop}%
\bibitem [{\citenamefont {Yue}\ and\ \citenamefont {Gaarde}(2020)}]{Yue_2020}%
  \BibitemOpen
  \bibfield  {author} {\bibinfo {author} {\bibfnamefont {L.}~\bibnamefont
  {Yue}}\ and\ \bibinfo {author} {\bibfnamefont {M.~B.}\ \bibnamefont
  {Gaarde}},\ }\bibfield  {title} {\bibinfo {title} {Imperfect recollisions in
  high-harmonic generation in solids},\ }\href
  {https://doi.org/10.1103/PhysRevLett.124.153204} {\bibfield  {journal}
  {\bibinfo  {journal} {Physical Review Letters}\ }\textbf {\bibinfo {volume}
  {124}},\ \bibinfo {pages} {153204} (\bibinfo {year} {2020})}\BibitemShut
  {NoStop}%
\bibitem [{\citenamefont {Garejev}\ \emph {et~al.}(2014)\citenamefont
  {Garejev}, \citenamefont {Gra{\v{z}}ulevi{\v{c}}i{\=u}t{\.e}}, \citenamefont
  {Majus}, \citenamefont {Tamo{\v{s}}auskas}, \citenamefont {Jukna},
  \citenamefont {Couairon},\ and\ \citenamefont {Dubietis}}]{Garejev_2014}%
  \BibitemOpen
  \bibfield  {author} {\bibinfo {author} {\bibfnamefont {N.}~\bibnamefont
  {Garejev}}, \bibinfo {author} {\bibfnamefont {I.}~\bibnamefont
  {Gra{\v{z}}ulevi{\v{c}}i{\=u}t{\.e}}}, \bibinfo {author} {\bibfnamefont
  {D.}~\bibnamefont {Majus}}, \bibinfo {author} {\bibfnamefont
  {G.}~\bibnamefont {Tamo{\v{s}}auskas}}, \bibinfo {author} {\bibfnamefont
  {V.}~\bibnamefont {Jukna}}, \bibinfo {author} {\bibfnamefont
  {A.}~\bibnamefont {Couairon}},\ and\ \bibinfo {author} {\bibfnamefont
  {A.}~\bibnamefont {Dubietis}},\ }\bibfield  {title} {\bibinfo {title}
  {Third-and fifth-harmonic generation in transparent solids with
  few-optical-cycle midinfrared pulses},\ }\href
  {https://doi.org/10.1103/PhysRevA.89.033846} {\bibfield  {journal} {\bibinfo
  {journal} {Physical Review A}\ }\textbf {\bibinfo {volume} {89}},\ \bibinfo
  {pages} {033846} (\bibinfo {year} {2014})}\BibitemShut {NoStop}%
\end{thebibliography}%
\end{document}